\documentclass[natbib]{emulateapj}
\usepackage{natbib}
\input epsf
\def\s2n{S^{\prime}/N}

\graphicspath{{figures/}}

\usepackage{amssymb,amsmath}
\def\bs{\boldsymbol}

\slugcomment{Submitted to ApJ, \today}
\shorttitle{Supernova Driving. IV. The Star Formation Rate of MCs}

\shortauthors{Padoan et al.}

\begin{document}
\title{Supernova Driving. IV. The Star Formation Rate of Molecular Clouds}

\author{Paolo Padoan,}
\affiliation{Institut de Ci\`{e}ncies del Cosmos, Universitat de Barcelona, IEEC-UB, Mart\'{i} i Franqu\`{e}s 1, E08028 Barcelona, Spain; ppadoan@icc.ub.edu}
\affiliation{ICREA, Pg. Llu\'{i}s Companys 23, 08010 Barcelona, Spain}
\author{Troels Haugb{\o}lle,}
\affiliation{Centre for Star and Planet Formation, Niels Bohr Institute and Natural History Museum of Denmark, University of Copenhagen, {\O}ster Voldgade 5-7, DK-1350 Copenhagen K, Denmark; haugboel@nbi.ku.dk}
\author{{\AA}ke Nordlund,}
\affiliation{Centre for Star and Planet Formation, Niels Bohr Institute and Natural History Museum of Denmark, University of Copenhagen, {\O}ster Voldgade 5-7, DK-1350 Copenhagen K, Denmark; aake@nbi.ku.dk}
\author{S{\o}ren Frimann}
\affiliation{Institut de Ci\`{e}ncies del Cosmos, Universitat de Barcelona, IEEC-UB, Mart\'{i} i Franqu\`{e}s 1, E08028 Barcelona, Spain; sfrimann@icc.ub.edu}

\begin{abstract}

We compute the star formation rate (SFR) in molecular clouds (MCs) that originate {\it ab initio} in a new, higher-resolution simulation of supernova-driven 
turbulence. Because of the large number of well-resolved clouds with self-consistent boundary and initial conditions, we obtain a large range of cloud physical 
parameters with realistic statistical distributions, an unprecedented sample of star-forming regions to test SFR models and to interpret observational surveys. 
We confirm the dependence of the SFR per free-fall time, $SFR_{\rm ff}$, on the virial parameter, $\alpha_{\rm vir}$, found in previous simulations, and 
compare a revised version of our turbulent fragmentation model with the numerical results. The dependences on Mach number, ${\cal M}$, gas to magnetic 
pressure ratio, $\beta$, and compressive to solenoidal power ratio, $\chi$ at fixed $\alpha_{\rm vir}$ are not well constrained, because of random scatter due 
to time and cloud-to-cloud variations in $SFR_{\rm ff}$. We find that $SFR_{\rm ff}$ in MCs can take any value in the range $0 \le SFR_{\rm ff} \lesssim 0.2$, 
and its probability distribution peaks at a value $SFR_{\rm ff}\approx 0.025$, consistent with observations. The values of $SFR_{\rm ff}$ and the scatter in the 
$SFR_{\rm ff}$--$\alpha_{\rm vir}$ relation are consistent with recent measurements in nearby MCs and in clouds near the Galactic center. Although not explicitly 
modeled by the theory, the scatter is consistent with the physical assumptions of our revised model and may also result in part from a lack of statistical equilibrium 
of the turbulence, due to the transient nature of MCs.

\end{abstract}


\section{Introduction}

Star formation in galaxies is ultimately regulated by the external supply of gas that can rapidly cool and settle at the high densities of star-forming
regions. The star formation rate (SFR) probed by the Kennicut-Schmidt relation \citep{Kennicutt98}, or by the Madau plot \citep{Madau+98}, 
is essentially determined by the cosmological  environment controlling the mass infall from the cosmic web onto the galaxies \citep[e.g.][]{Dekel+09,Forbes+14}
and the radial transport in the galactic disk from low density gas at large radii to high density gas at smaller radii. 
The dark matter component of this boundary condition of galaxies is well understood in the standard $\Lambda$CDM model, while the structure of the baryonic 
component of the infall has been studied with large dynamic range simulations including gas dynamics. However, galaxies can achieve a global SFR in approximate 
balance with the cosmological infall through a variety of paths, leading to or requiring different galaxy morphologies and different dynamical and chemical evolutions. 
For example, the star formation may be smoothly distributed in space or time, or may occur in rapid bursts within massive clouds. The detailed mode of star formation 
is only partly dependent on the large-scale boundary conditions; it is primarily controlled by disk dynamics (e.g. gravitational instability and radial transport) and by 
the physics of star formation, meaning the specific processes responsible for the evolution and fragmentation of cold interstellar medium (ISM) clouds. The theoretical 
modeling of such processes should result in a physical law that predicts the SFR as a function of star-forming cloud parameters. The formation and evolution of galaxies 
cannot be fully understood until such a universal SFR law is revealed. 

Despite recent progress in modeling the effect of supersonic turbulence on the fragmentation of star-forming clouds, the development of large dynamic range simulations
of star formation, and the ever growing number and size of galactic and extragalactic surveys of star forming regions, SFR laws remain very hard to test. The SFR in MCs
is difficult to measure, with different methods often yielding very different values. The cloud physical parameters on which the theoretical models depend (for example the 
ratio of gas and magnetic pressures, or the ratio of compressive to solenoidal power of the velocity field) are difficult to measure as well. Furthermore, recent
compilations of MC properties including their SFR have shown that the SFR has only a weak dependence on cloud parameters and a very large scatter 
\citep[e.g.][]{Murray11,Evans+14,Vuti+16}, in apparent contradiction of most theoretical models. Besides the very large scatter, the observed SFRs 
also tend to be lower than both numerical estimates and theoretical predictions. 

On the other hand, idealized numerical simulations of supersonic turbulence have been used to test theoretical SFR laws with some success \citep{Padoan+Nordlund11sfr,
Federrath+Klessen12,Padoan+12sfr}. They have also yielded best-fit values of the free parameters of the theoretical models. These simulations tend to give larger 
SFR than the observations, even when the turbulence is well resolved and relatively strong magnetic fields are included. More importantly, they do not yield the large 
scatter in SFR, for fixed physical parameters, found in the observations. Thus, while our theoretical understanding of the star formation process has improved 
significantly with the interplay of theory and simulations, our ability to reproduce the observations remains limited, casting doubts upon the 
theoretical scenario of turbulent fragmentation. Nevertheless, we believe the discrepancies with the observations are mainly the consequence of the limitations of the 
simulations. More realistic simulations should reproduce the observations and guide a revision of the theoretical SFR models, while retaining the key idea that star-forming 
clouds are fragmented by supersonic turbulence. 

The main limitation of the idealized simulations used to test the SFR models is that they capture the fundamental physics of supersonic MHD 
turbulence, without describing a full cloud, its realistic boundary conditions and its realistic evolution including the cloud formation and dispersion processes. The simulations
adopt periodic boundary conditions, so they are interpreted as a characteristic piece of a MC, and they are started from idealized initial conditions, so they must be evolved 
for several dynamical times to pursue a statistical steady state, before gravity is introduced. Because of the need to first develop the turbulence, gravity has to be suddenly 
included at a later time, so even the initial evolution under the effect of gravity is not entirely realistic; the simulation has to be run for at least one free-fall time with gravity,
before the SFR is measured. The SFR is initially very low, when gravity is first included, and then gradually increases, in the best case (but not always) becoming 
approximately constant for at least a few free-fall times. It is this late time, approximately (or `hopefully') constant SFR that is usually measured in the simulations, to avoid
the initial transient phase that reflects the imprint of the numerical setup. Thus, the SFR derived in this way from the simulations does not reflect the initial time variations,
because it does not record the low SFR values of the initial transient phase. 

Real MCs also have an initial phase when they are first assembled and their SFR may be very small. They may also become more quiescent after exhausting 
most of their dense gas, or reduce their SFR rapidly during a period of cloud dispersion, towards the end of their lifetime. But this time evolution of MCs and 
variability of their SFR cannot be described with the idealized simulations mentioned above. Thus, it is to be expected that only the highest SFR values of MCs
are reproduced by the simulations, while lower values are neglected because their frequency cannot be estimated with the idealized numerical setups. Furthermore,
the physical parameters that controls the SFR, such as Mach number, virial parameter and gas to magnetic pressure ratio, are usually averaged over the whole
computational domain, which is not the same as measuring them for a whole MC. 

The alternative approach of simulating a whole cloud from the moment of its formation must rely on ad hoc and idealized initial conditions, where velocity,
density and magnetic fields can only mimic and grossly misrepresent the turbulence in real MCs. Because gravity is present from the beginning, the star formation 
process starts rapidly, and the artificial initial conditions leave a strong imprint on star formation. SFR values derived in this way cannot be used to faithfully sample 
the variations of the SFR in real MCs.

The only way to model the range of variations of the SFR in MCs is to adopt a numerical setup where MCs are formed {\it ab initio}, meaning without ad hoc initial
or boundary conditions. This can only be achieved by simulating a volume much larger than the size of a single MC \citep[e.g.][]{Dobbs+Pringle13,Bonnell+13,Dobbs15}, 
which is expensive, but well worth the cost, because a large volume yields a large population of MCs, all emerging from initial and boundary conditions with realistic probability 
distributions. With a single large scale run, the SFR evolution can be followed over time in each cloud, generating realistic variations in time and from cloud to cloud as well. 
Very low SFR can then be realized as well, and both average values and scatter of the SFR as a function of physical parameters can be compared with the observations and 
with theoretical models, which is the goal of the present work.

Such a large scale simulation was already presented in our previous three papers of this series \citep[][Paper I, Paper II and Paper III hereafter]{Padoan+16SN_I,
Pan+16,Padoan+16SN_III}, where we studied the properties of the ISM turbulence driven by supernova (SN) explosions. The properties of MCs formed 
self-consistently in the turbulent ISM of the simulation were found to agree with those of real MCs from the $^{12}$CO FCRAO Outer Galaxy Survey
\citep{Heyer+98,Heyer+01}. In this work, we adopt the same numerical setup as a `laboratory' to study star formation in MCs, by increasing the spatial resolution 
of the simulation and including accreting sink particles to describe the formation of massive stars. In this new simulation, the turbulence is driven by 
the explosion of massive stars whose formation is resolved, so both the timing and position of the SNe, hence their effect on MCs \citep{Iffrig+Hennebelle15}, is 
now much more realistic than in the previous simulation (or any simulation of SN-driven turbulence to date), where the SNe where generated randomly in space and time. 
The main limitation of this setup is the lack of even larger scales, as our computational domain is a cubic volume of 250 pc size with periodic boundary conditions (hence 
fixed total mass), neglecting differential rotation and the vertical gravitational potential of the galactic disk. However, such limitation is unlikely to affect directly the 
SFR within individual MCs.

\begin{figure}[t]
\includegraphics[width=\columnwidth]{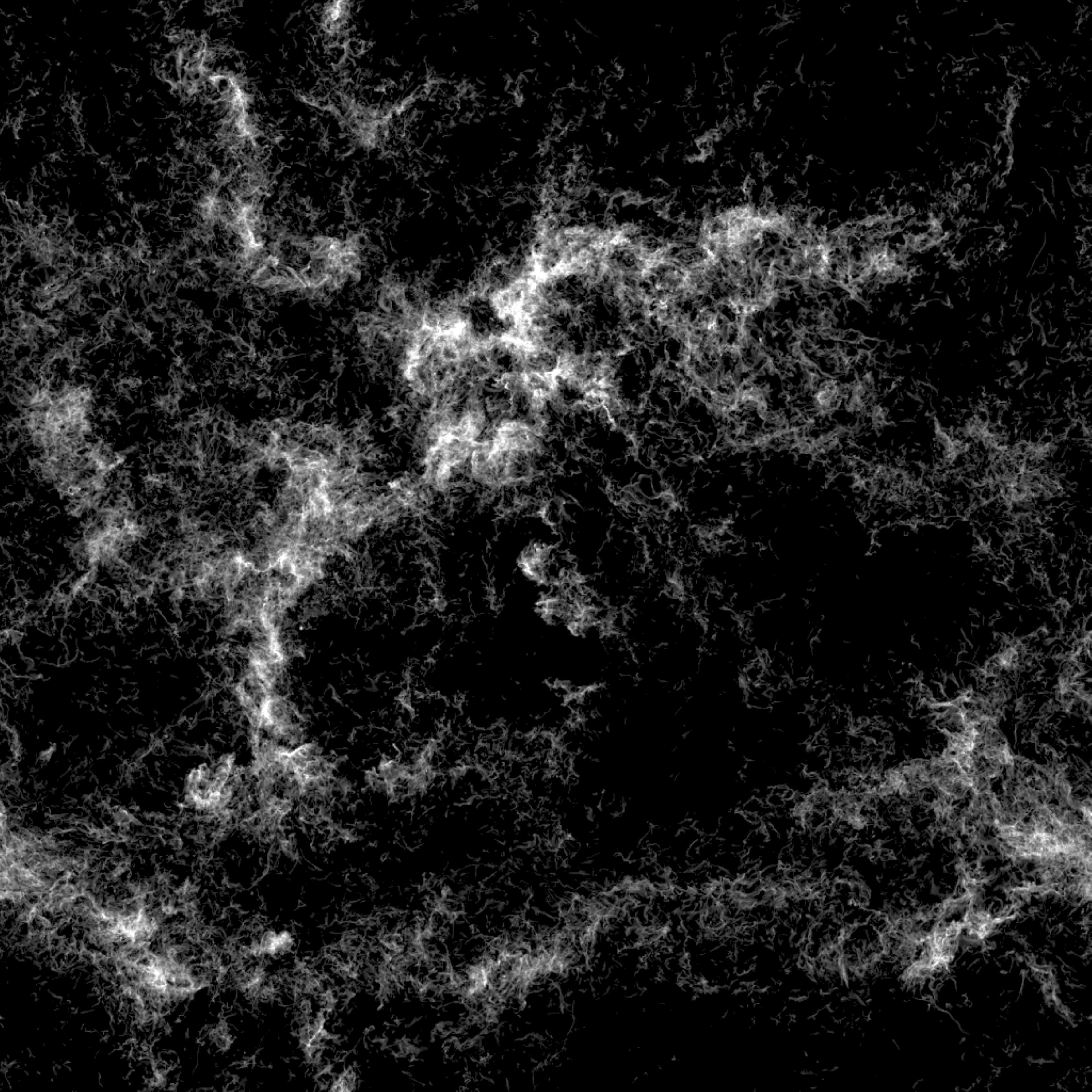}
\includegraphics[width=\columnwidth]{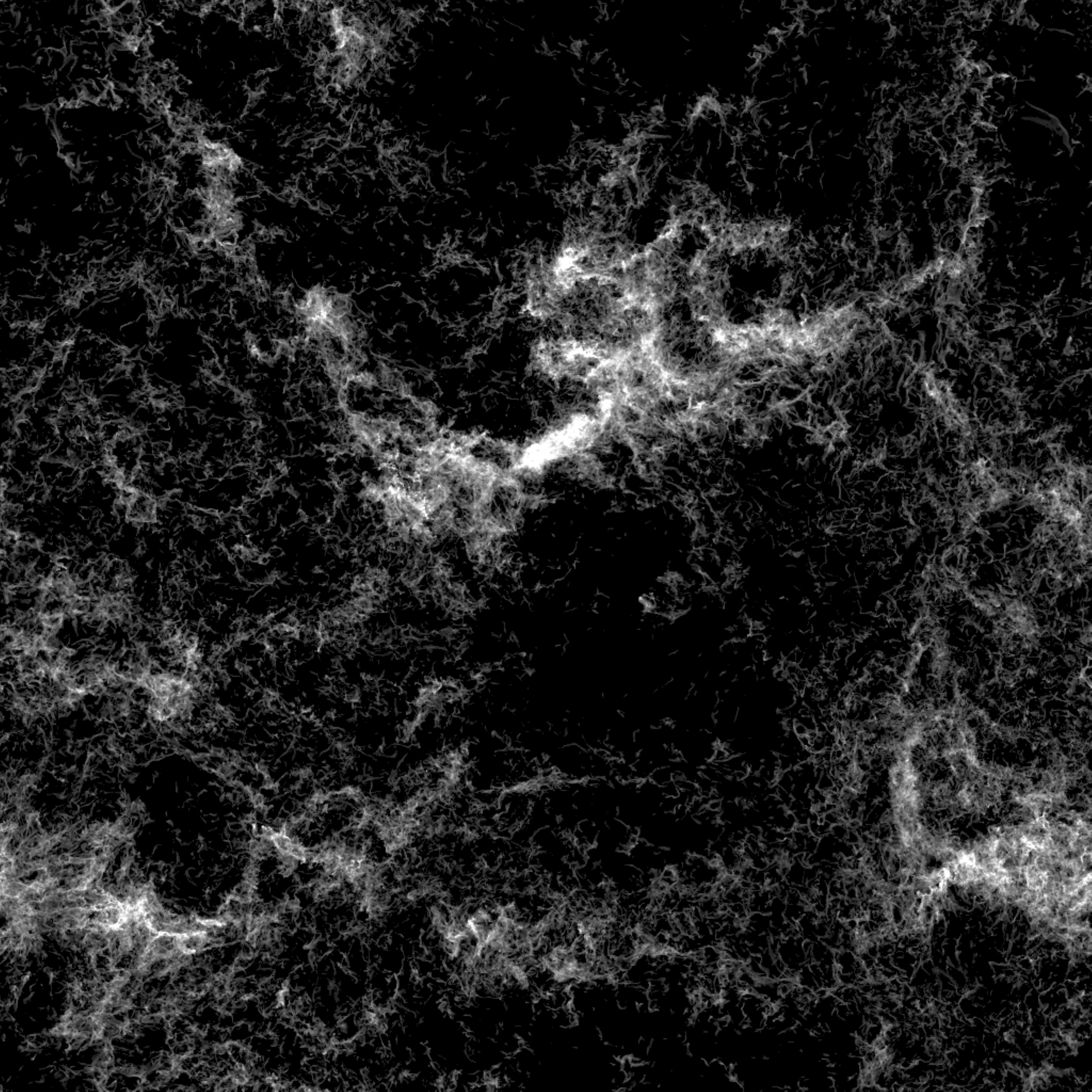}
\caption[]{Square root of the projected density of the whole 250 pc computational domain, integrated along the direction of the mean magnetic field and computed at 
$t=70.0$ Myr (upper panel) and $t=74.7$ Myr (lower panel), corresponding to 14.6 Myr and 19.3 Myr after the inclusion of self-gravity, respectively. The data has been 
extracted at a resolution of 0.24 pc. The grayscale intensity starts at a column density of approximately 1.0 M$_{\odot}$pc$^{-2}$ and saturates at a maximum column 
density of 149 M$_{\odot}$pc$^{-2}$. The projected density is computed using only cells with a number density above the threshold $n_{\rm H,min}=200$ cm$^{-3}$, 
to illustrate the mass distribution corresponding to our lower-density MC catalog.}
\label{images_gas}
\end{figure}
\begin{figure}[t]
\includegraphics[width=\columnwidth]{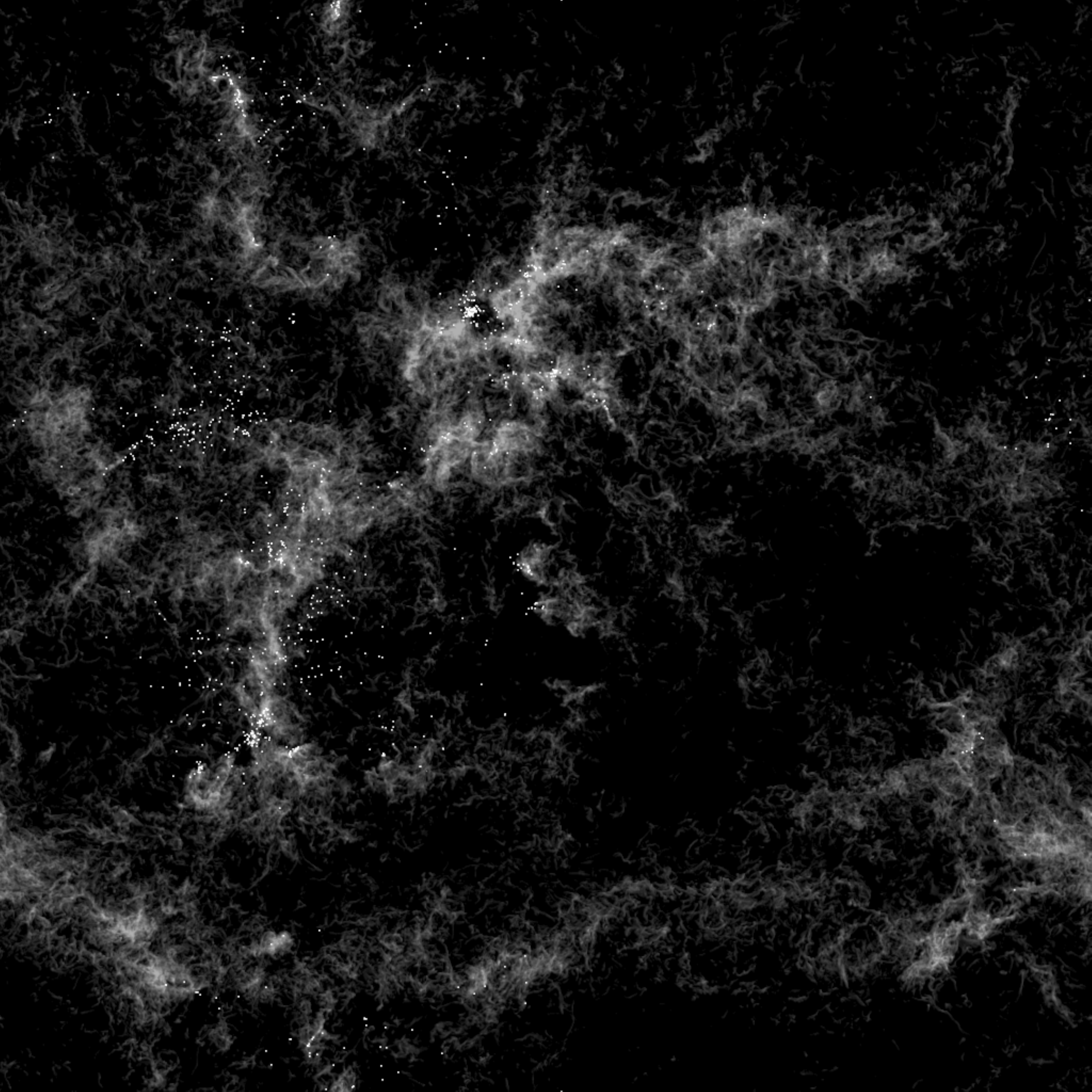}
\includegraphics[width=\columnwidth]{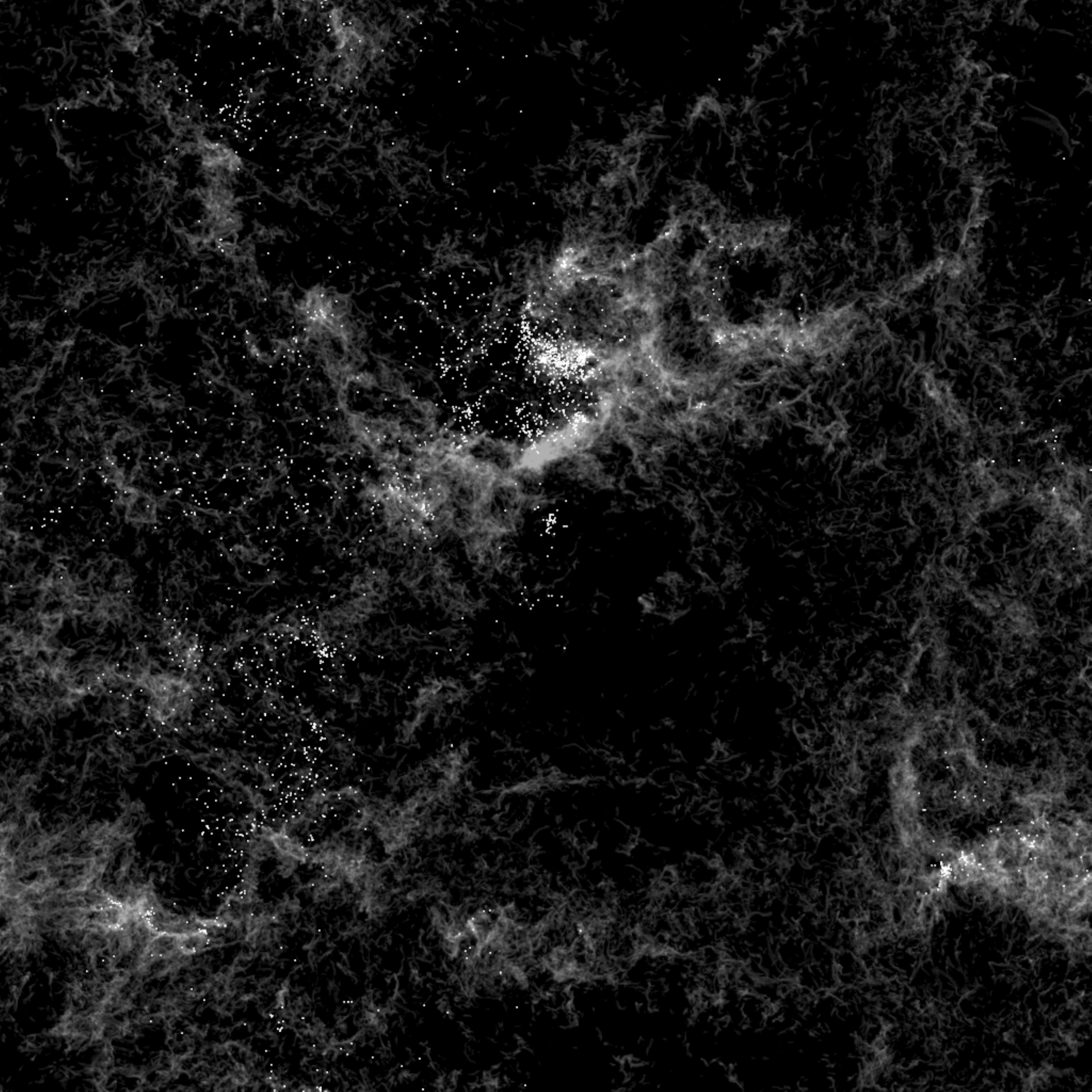}
\caption[]{The same projected-density fields as in Figure \ref{images_gas}, but with a compressed grayscale intensity range, making the saturation value of
149 M$_{\odot}$pc$^{-2}$ darker, thus allowing the sink particles to stand out as dots of the highest intensity value. All sink particles with mass larger than 1 M$_{\odot}$
are shown (the largest mass is 117.1 M$_{\odot}$), which amount to 1408 sinks in the upper panel and 2888 sinks in the lower panel. One can clearly see (young)
sinks inside the densest filaments, as well as many (older) sinks that have already left their parent clouds, including clusters that have cleared their 
surrounding gas thanks to SN explosions of their most massive members.}
\label{images_gas_sink}
\end{figure}

As in previous works, we normalize the SFR by the ratio of the total mass of the system, $M$, and the free-fall time of its mean density $\rho_0$, 
$t_{\rm ff}=(3\pi/32G\rho_0)^{1/2}$, to obtain a non-dimensional SFR referred to as ``SFR per free-fall time", $SFR_{\rm ff}\equiv -dM/dt / (M/t_{\rm ff})$, 
first introduced by \citet{Krumholz+McKee05sfr}. In \citet{Padoan+12sfr}, using a large set of AMR simulations, we derived a SFR per free-fall time that 
depended only on the virial parameter, $\alpha_{\rm vir}$ (the ratio of kinetic turbulent energy and gravitational energy --see definition in Section \ref{sect_mcs}), 
$SFR_{\rm ff,P12} = 0.5 \, \exp(-1.38 \, \alpha_{\rm vir}^{1/2})$ (see Section \ref{sec_model_pred}).
Because both the virial parameter and the SFR were derived as global values for the whole computational box, this result is suitable for the 
development of subgrid models of stellar feedbacks in galaxy formation simulations \cite[e.g.][]{Semenov+16}. In this work, we focus on
the dependence of the SFR on the properties on individual clouds, in order to test the theoretical predictions over a large statistical sample and
to compare with observational estimates of the SFR in real MCs. By applying a revision of our turbulent fragmentation model
to the physical parameters of the clouds extracted from the simulation, we derive a new empirical relation between the SFR per free-fall time and
the virial parameter that applies to individual clouds, $SFR_{\rm ff,\alpha} = 0.4 \, \exp(-1.6 \, \alpha_{\rm vir,e}^{1/2})$ (see Section \ref{sec_model_pred}), 
where $\alpha_{\rm vir,e}$ is the effective virial parameter of individual clouds (see Section \ref{sect_mcs}), rather than its estimated value over 
the whole computational domain. For reference, we show this new empirical law as a dashed line in several figures throughout the paper (figures \ref{sfr_alpha}, 
\ref{sfr_model}, \ref{sfr_alpha_model}, \ref{sfr_alpha_evans}), even preceding its derivation in Section \ref{sec_model_pred}.     

The paper is organized as follows. In the Section 2, we describe the numerical setup, and in Section 3 we define and compute the physical parameters of MCs 
selected from the simulation. The cloud SFR is studied in Section 4, where both average values and scatter are found to be realistic. We then summarize the 
turbulent fragmentation model of the SFR and propose a slight revision in Section 5. In the same section, we also apply our revised SFR model to the clouds from 
the simulation. On average, we find good agreement between the SFR from the revised model and the SFR measured directly from the simulation, while the scatter 
in the model is too small. We discuss the origin of the scatter arguing that it is actually predicted, though not accounted for, by the model. In Section 6 we compare 
our results with SFR estimates in real MCs, and in Section 7 we summarize our work and list the most important conclusions.

\section{Simulation}

This work is based on a new MHD simulation of SN-driven turbulence in a large ISM volume, carried out with the Ramses AMR code \citep{Teyssier02,Fromang+06,Teyssier07}. 
The numerical method and setup are discussed extensively in Paper I and are only briefly summarized here. We simulate 
a cubic region of size $L_{\rm box}=250$ pc, with periodic boundary conditions and a total mass of $M_{\rm box}=1.9\times 10^6$ $M_{\odot}$. The initial condition for this 
simulation is the final snapshot of our previous SN-driven simulation (see Paper I) before the introduction of self-gravity, at $t=45$ Myr. That simulation was 
started with zero velocity, a uniform density $n_{\rm H,0}=5$ cm$^{-3}$, a uniform temperature $T_0=10^4$ K, and a uniform magnetic field $B_0=4.6$ $\mu$G, 
later amplified by the turbulence to an rms value of 7.2 $\mu$G and an average of $|\bs B|$ of 6.0 $\mu$G, consistent with the value of $6.0 \pm 1.8$ $\mu$G 
derived from the `Millennium Arecibo 21-cm Absorption-Line Survey' by \citet{Heiles+Troland05}. SN explosions were randomly distributed in space and time 
(see discussion in Paper I in support of this choice), with a rate of 6.25 SNe Myr$^{-1}$. The resolution was $dx=0.24$ pc, with a $128^3$ root grid and three 
AMR levels. Details about the numerical setup and the implementation of random SN driving, tracer particles, and parametrized heating and cooling can be 
found in Paper I. In the following, we focus on the description of the new features of the current simulation.

\begin{figure}[t]
\includegraphics[width=\columnwidth]{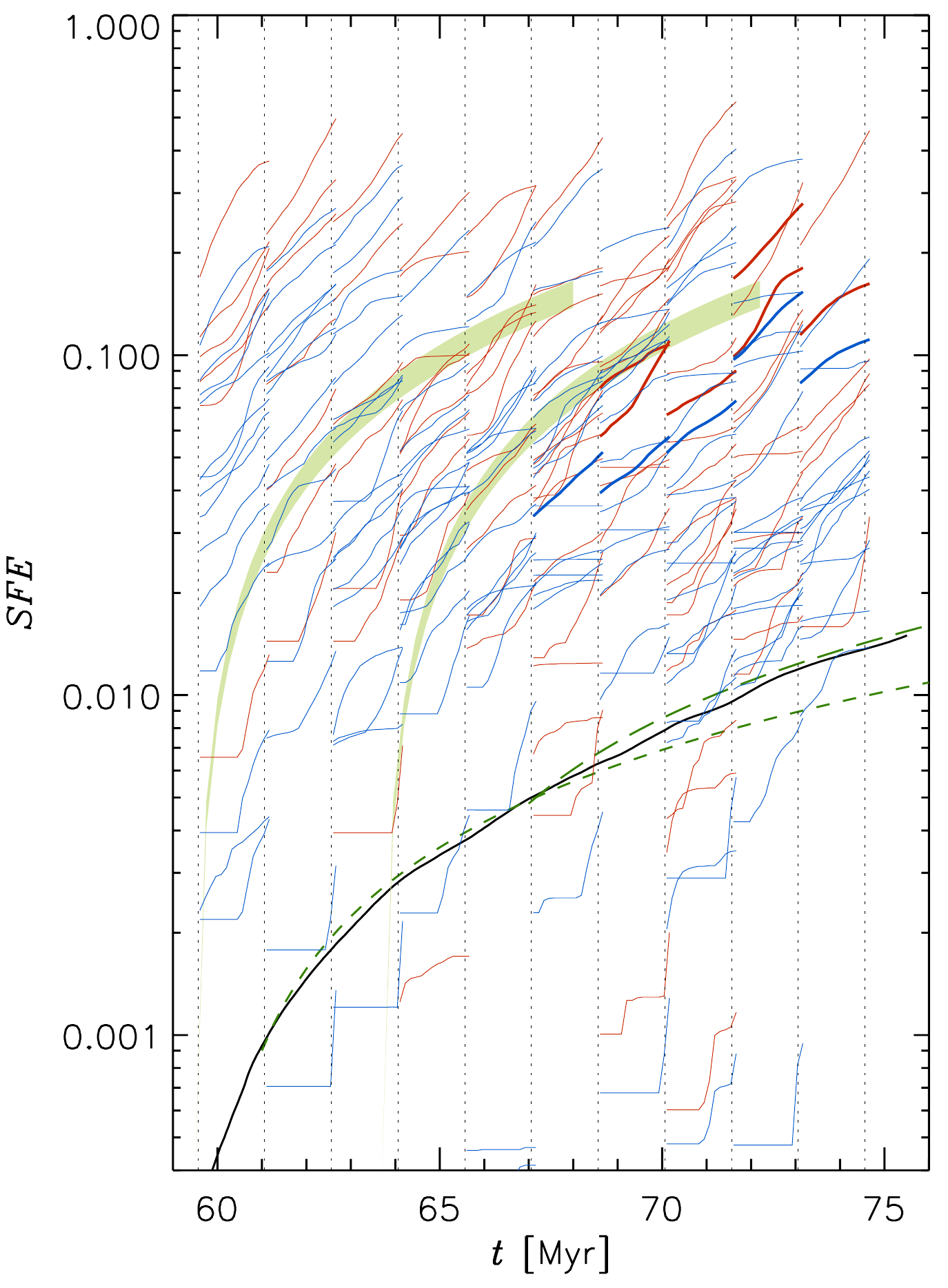}
\caption[]{Star formation efficiency versus time for the whole computational volume (black thick solid line), and for individual MCs from the simulation, selected 
with $n_{\rm H,min}=200$ cm$^{-3}$ (blue thin lines) and $n_{\rm H,min}=400$ cm$^{-3}$ (red thin lines). The thin line of each cloud covers 1.68 Myr, the time
interval during which the clouds are followed (notice that the SFE for five MCs with $M_{\rm cl} > 2\times10^4$ M$_{\odot}$ ($n_{\rm H,min}=200$ cm$^{-3}$) 
and for six MCs with $M_{\rm cl} > 10^4$ M$_{\odot}$ ($n_{\rm H,min}=400$ cm$^{-3}$) has been plotted with thicker lines). The vertical dotted lines mark the 
times of the 10 simulation snapshots where the MCs were selected, at intervals of 1.5 Myr. The short-dashed and long-dashed lines show $SFE$ versus time 
for depletion times $t_{\rm dep}=1.5$ Gyr starting at $t=61$ Myr and $t_{\rm dep}=0.8$ Gyr starting at $t=67$ Myr, respectively. The two green shaded areas 
show two examples of $SFE$ versus time for $t_{\rm dep}=0.05$ Gyr, characteristic of the MCs from the simulation, starting at the first and fourth cloud selection 
times. This cloud depletion time is evaluated as $1/ \langle (SFR_{\rm ff}/t_{\rm ff})\rangle$ averaged over all clouds with non-zero SFR.}
\label{sfe_time}
\end{figure}

Because we wish to resolve the formation of individual massive stars, we increase significantly the resolution
relative to our previous run. As we restart at $t=45$ Myr, we continue to run without self-gravity, and increase the root grid size to $512^3$ cells,
besides adding four AMR levels to reach a minimum cell size of $dx=0.03$ pc.  We also initialize 250 million passively advected tracer particles,
each representing a fluid element with a characteristic mass of approximately 0.008 $M_{\odot}$. The tracers record all the hydrodynamic variables,
and are tagged once they accrete onto a sink particle, so the star formation process can be entirely followed through the Lagrangian point of view of the tracers.

One of the goals of the new simulation is to describe the SN driving with a precise knowledge of the time and location of each SN, by resolving the formation
of all massive stars. To reduce the role of the `artificial' randomly generated SNe, we gradually decrease their rate as the rate of the realistically generated 
SNe increases, so we start by reducing the random SN rate by a factor of two. We continue the simulation for 10.5 Myr, with a random SN rate of only 
3.12 SNe Myr$^{-1}$, with the increased resolution, and with the new set of tracer particles. At $t=55.5$ Myr we include self-gravity and add two more 
AMR levels, reaching a maximum resolution of $dx=0.0076$ pc. This resolution yields a complete stellar initial mass function (IMF) down to a stellar
mass of order 5-10 M$_{\odot}$ (though significantly lower-mass sink particles are also formed). The IMF is approximately a power law, with a slope only
slightly steeper than the observed Saltpeter's value, thus the relative number of SNe, resulting from the explosion of the massive stars formed
in the simulation, has a realistic mass and time dependence (besides a realistic spatial istribution).

To follow the collapse of prestellar cores, sink particles are created in cells where the gas density is larger than $10^6$ cm$^{-3}$
(approximately 10 times larger than the largest density reached by the turbulence without self-gravity), if the following additional conditions are met at the 
cell location: i) The gravitational potential has a local minimum value, ii) the three-dimensional velocity divergence is negative, and iii) no other previously 
created sink particle is present within an exclusion radius, $r_{\rm excl}$ ($r_{\rm excl}=16 dx=0.12$ pc in this simulation). These conditions are similar to
those in \citet{Federrath+10sinks}. We have verified that they avoid the creation of spurious sink particles in regions where the gas is not collapsing. 
An extensive presentation of our sink particle implementation in Ramses can be found in Haugb{\o}lle et al. (2017). 
 
We plan to continue this simulation for $\sim 50-100$ Myr with self-gravity and sink particles, in order to reach a fully self-consistent solution, where the ISM
turbulence and the star formation process are driven entirely by the explosion of massive stars formed in the simulation, including the least massive ones,
of approximately 7.5 M$_{\odot}$, that have a lifetime of nearly 50 Myr \citep{Schaller+92}\footnote{Neglecting even longer lifetimes for core-collapse SNe from the interaction
of intermediate-mass binaries \citep{Zapartas+17}}. So far, we have reached $t\approx 75.5$ Myr, that is 20 Myr with self-gravity, yielding more than
6000 sink particles, a sufficiently long time for the purpose of this paper, which is to study the SFR in MCs. Even at the current stage, this is a challenging 
computational project that has already used approximately 20 million core hours on the Pleiades supercomputer at NASA/Ames. 

The projected density from two snapshots of the simulation taken at 14.6 and 19.3 Myr after the inclusion of self-gravity are shown in Figure \ref{images_gas}.
The gas distribution is highly filamentary on all scales and all densities, with large voids created by the explosions of multiple SNe. In Figure \ref{images_gas_sink},
the grayscale intensity range has been compressed, and all the sink particles with mass larger than 1 M$_{\odot}$ (1408 and 2888 sink particles in the upper and
lower panel respecitely) have been overplotted. Young sinks are found inside the densest filaments, while older ones have already left their parent clouds. Some
dense clusters have also cleared their surrounding gas thanks to SN explosions of their most massive members.

The large number of MCs generated by this simulation, in combination with several hundreds of SNe from resolved massive stars and several thousands
of sink particles, make up an unprecedented numerical sample to study the interaction of SNe with their parent clouds and to investigate the amount of clustering 
in space and time of SNe of different masses. These studies are deferred to a separate paper. The analysis of the global SFR under this fully self-consistent
SN driving will also be addressed elsewhere, once the simulation will be closer to the final goal of $\sim 50-100$ Myr with self-gravity and sink particles. 
Here, it is worth mentioning that, while producing realistic SFR values within MCs, this simulation has currently a global SFR corresponding to a 
gas depletion time of the order of 1 Gyr, consistent with global galactic values \citep{Bigiel+11}. Figure \ref{sfe_time} shows the global star formation
efficiency, $SFE$, versus time (solid line), where
\begin{equation}
SFE(t)\equiv M_{\rm s}(t)/M_{\rm box}
\label{eq_sfe_tot}
\end{equation}
and $M_{\rm s}$ the total mass in sink particles. The depletion time, $t_{\rm dep}$, is defined as:
\begin{equation}
t_{\rm dep}\equiv M_{\rm box} / (dM_{\rm s}(t)/dt)=1 / (dSFE(t)/dt).
\label{eq_t_dep}
\end{equation}
In Figure \ref{sfe_time}, the short-dashed and long-dashed lines show $SFE$ versus time for depletion times $t_{\rm dep}=1.5$ and 0.8 Gyr, respectively. 
One can see that the global SFR corresponds approximately to these values of $t_{\rm dep}$ in the approximate time intervals 61-67 Myr and  
67-74 Myr. At the same time, the clouds selected in the simulation as described in the following section exhibits a much steeper time dependence of 
their local $SFE$, as shown by the short thin lines in Figure \ref{sfe_time}, and by the shaded areas illustrating the average cloud depletion time of
51 Myr. This is the first time that a value of $t_{\rm dep}$ characteristic of global galactic values is derived in a simulation where both the star formation 
and its feedback are resolved, for each individual massive stars, instead of being imposed with subgrid-scale models.

\section{Molecular Cloud Parameters} \label{sect_mcs}

The simulation does not model the formation of ISM molecules, so we define the clouds simply as connected regions above 
a threshold density, $n_{\rm H,min}$, and refer to them as MCs. To keep track of the effect of the value of $n_{\rm H,min}$, we 
consider two values, $n_{\rm H,min}=200$ cm$^{-3}$ and  $n_{\rm H,min}=400$ cm$^{-3}$. The MC search is carried out at a 
uniform resolution of $512^3$ cells, that is a spatial resolution of 0.49 pc, but the cloud properties are computed using all the tracer 
particles identified within each cloud. Because the tracers record all the hydrodynamical variables interpolated at their position, and 
due to the very large number of tracers in high density regions, MC properties are derived with the hydrodynamical variables sampled
at the highest local spatial resolution of the AMR grid, up to the highest resolution of 0.0076 pc in the densest regions. 

MCs are selected from 10 snapshots, at equal intervals of 1.5 Myr, with the first snapshot at 4.0 Myr after the inclusion of 
self-gravity in the simulation. Only clouds satisfying the following three conditions are retained: 1) the cloud mass is 
$M_{\rm cl} > 1000$ M$_{\odot}$, 2) the rms velocity is $\sigma_{\rm v}< 4$ km/s (to avoid MCs too strongly affected by recent 
nearby SNe\footnote{Nearby SNe may disperse the MC (that is addressed by the third condition for cloud selection), or only accelerate 
a small portion of the cloud, causing a strong but temporary increase of the cloud rms velocity, with very little influence on its SFR.}),
3) the cloud does not disperse during the next 1.5 Myr, meaning that it is not doubling its effective size in that time interval (to avoid MCs 
whose properties are evolving too rapidly). With these conditions, the number of MCs is reduced from 391 to 313, 203 clouds with 
$n_{\rm H,min}=200$ cm$^{-3}$ and 110 clouds with $n_{\rm H,min}=400$ cm$^{-3}$. The conditions allow for a better comparison
with the SFR model that does not account for transient processes like SN feedback or cloud dispersal. It may yield a sample more 
suitable for the comparison with the observations as well, because MCs strongly affected by SNe, or in the process of being dispersed,
may also suffer from a strong feedback by HII regions, which is not modeled in the simulation. 
 
In the following, we will compute for each cloud the three non-dimensional parameters, $\alpha_{\rm vir}$, ${\cal M}$ and $\beta$, 
expressing the ratios of turbulent, gravitational, thermal and magnetic energies. The virial parameter, 
$\alpha_{\rm vir}$, estimates the ratio of turbulent and gravitational energies in a spherical cloud of 
radius $R_{\rm cl}$, mass $M_{\rm cl}$ and one-dimensional velocity dispersion $\sigma_{\rm v}$ \citep{Bertoldi+McKee92}:
\begin{equation}
\alpha_{\rm vir} \equiv {5 \sigma^2_{\rm v} R_{\rm cl} \over{G M_{\rm cl}}} = {40\over 3\pi^2}\left({t_{\rm ff}\over t_{\rm dyn}}\right)^2 \sim {2 E_{\rm k}\over{E_{\rm g}}},
\label{eq_alpha}
\end{equation}
where the dynamical time is defined as:
\begin{equation}
t_{\rm dyn} \equiv R_{\rm cl}/\sigma_{\rm v,3D}.
\label{eq_tdyn}
\end{equation}
The last equality in (\ref{eq_alpha}) is exact in the case of an idealized spherical cloud of uniform density. For more realistic cloud mass distributions,
the virial parameter is only an approximation of the ratio of kinetic and gravitational energies. The rms Mach number is the ratio
of the three-dimensional rms velocity and the sound speed, $c_{\rm s}$:
\begin{equation}
{\cal M} \equiv \sigma_{\rm v,3D} / c_{\rm s},
\label{eq_mach}
\end{equation}
and $\beta$ is the ratio of gas to magnetic pressure:
\begin{equation}
\beta \equiv P_{\rm g} / P_{\rm m}=2\, \gamma^{-1} ({\cal M}_{\rm A}/{\cal M})^2,
\label{eq_beta}
\end{equation}
where ${\cal M}_{\rm A}$ is the rms Alfv\'{e}nic Mach number, ${\cal M}_{\rm A}=\sigma_{\rm v,3D}/v_{\rm A}$, with $v_{\rm A}$ the Alfv\'{e}n velocity,
$\gamma$ is the adiabatic index, and we have used the adiabatic sound speed, $c_{\rm s}=\sqrt{\gamma P_{\rm g} / \rho}$.

\begin{figure*}[t]
\centering
\includegraphics[width=\columnwidth]{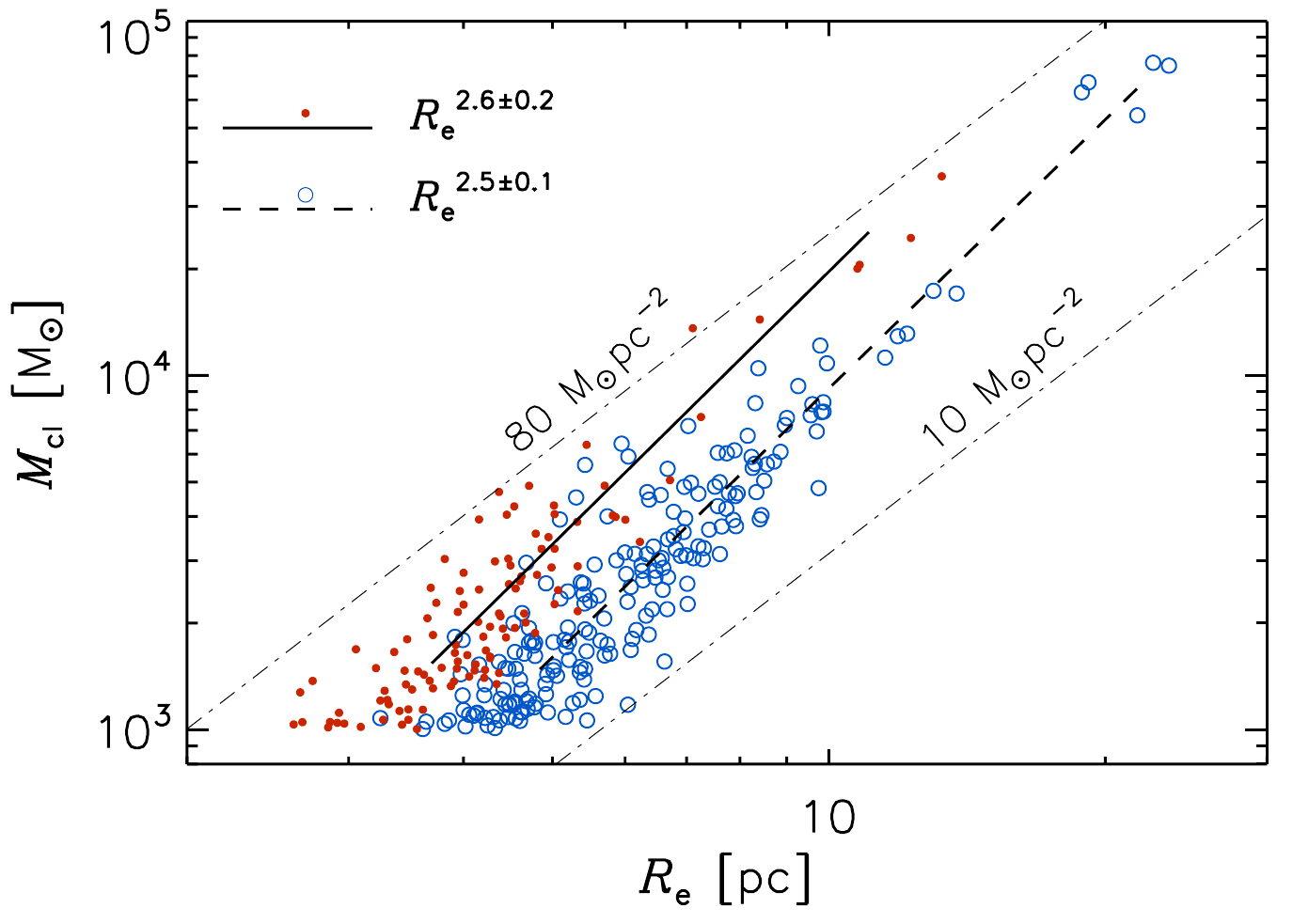}
\includegraphics[width=\columnwidth]{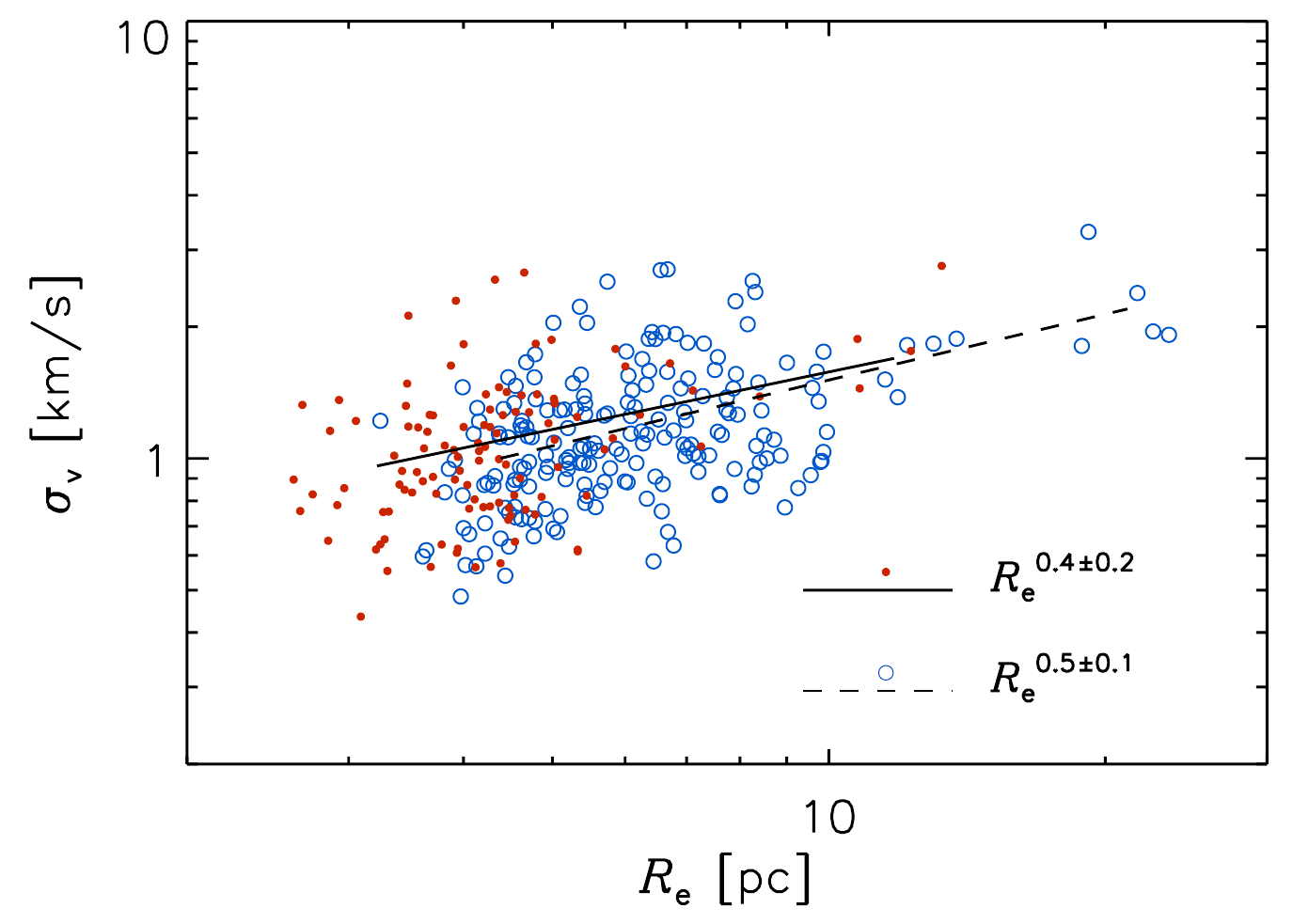}
\includegraphics[width=\columnwidth]{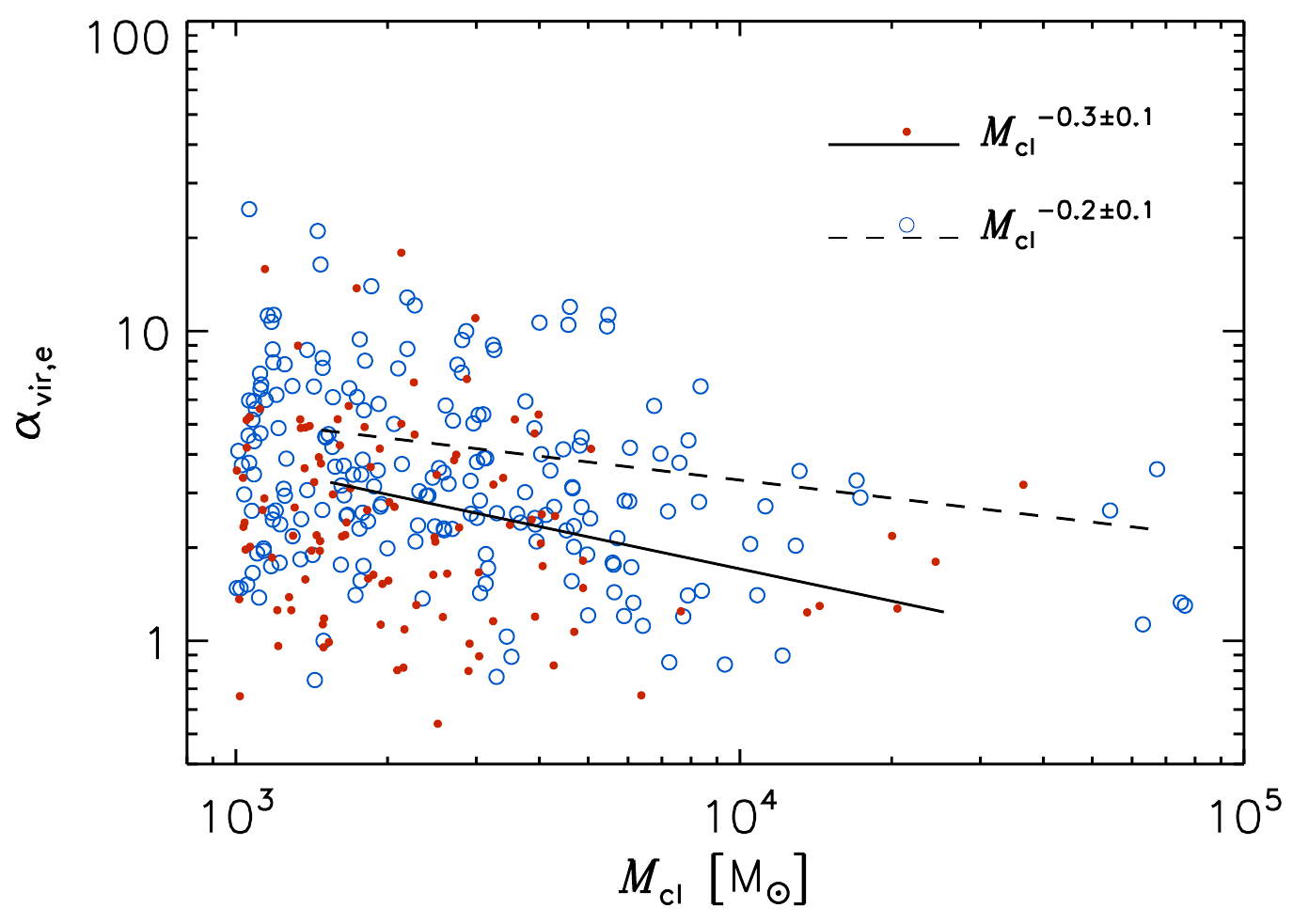}
\includegraphics[width=\columnwidth]{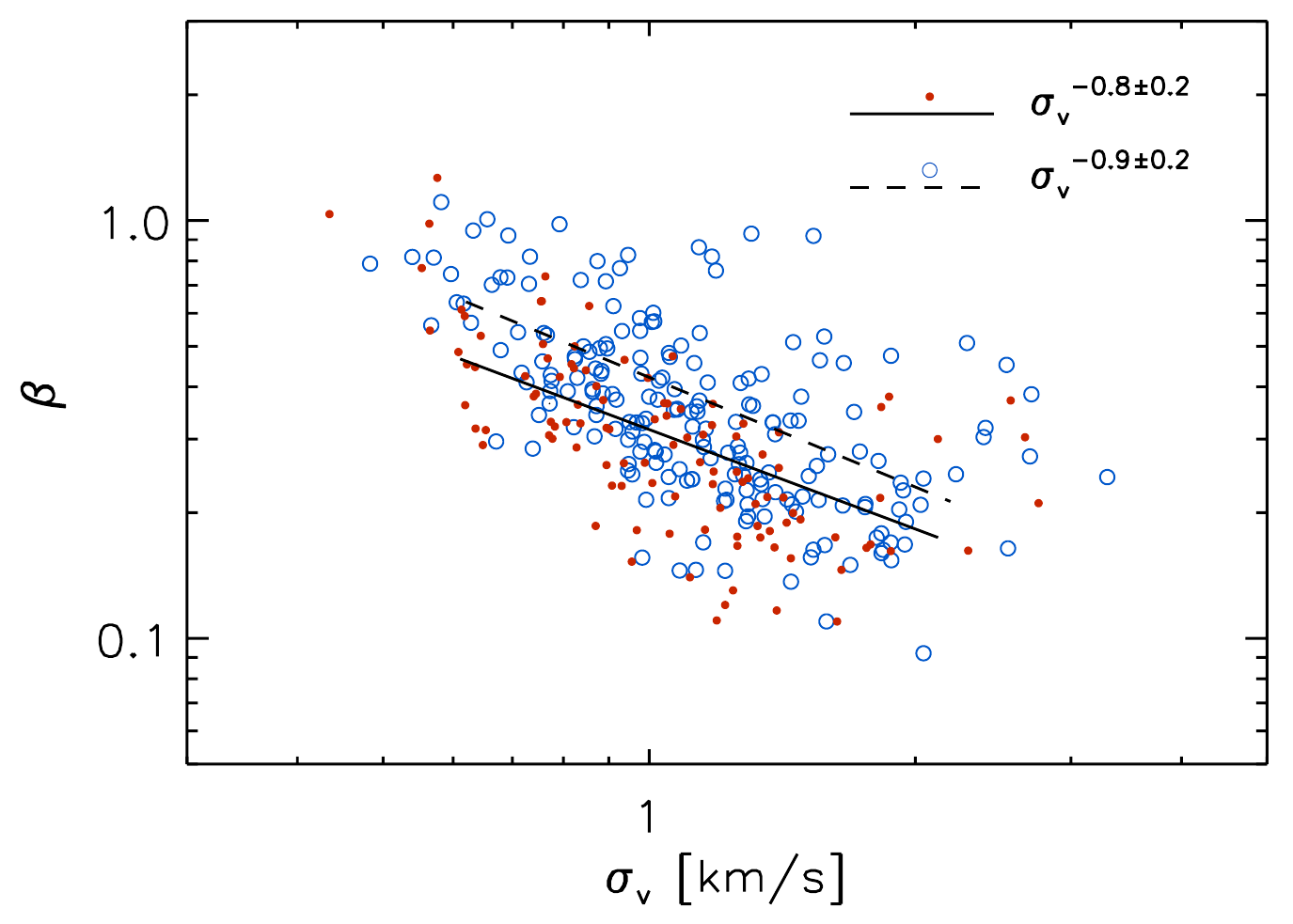}
\caption[]{Properties of MCs selected from the simulation with density thresholds $n_{\rm H,min}=200$ cm$^{-3}$ (empty circles) and 
$n_{\rm H,min}=400$ cm$^{-3}$ (filled circles). The solid and dashed lines are the power-law fits to the mean values of $M_{\rm cl}$
and $\sigma_{\rm v}$ averaged within logarithmic intervals of $R_{\rm e}$ (upper panels), and the values of $\alpha_{\rm vir,e}$ and $\beta$
averaged inside logarithmic intervals of $M_{\rm cl}$ (lower left panel) and $\sigma_{\rm v}$ (lower right panel).}
\label{mc_properties}
\end{figure*}

To estimate the non-dimensional parameters for MCs, we need to compute the cloud mass, $M_{\rm cl}$, radius, $R_{\rm cl}$, 
velocity dispersion, $\sigma_{\rm v}$, sound speed, $c_{\rm s}$, magnetic pressure, $P_{\rm m}$ and thermal pressure $P_{\rm g}$. 
The cloud radius is defined as the equivalent radius of the circle with area equal to the cloud projected (along a randomly chosen axis
direction) area, $A_{\rm cl}$,
\begin{equation}
R_{\rm e}\equiv\sqrt{A_{\rm cl}/\pi}.
\label{eq_Re}
\end{equation}
The cloud rms velocity is defined as the density-weighted one-dimensional rms velocity, and is computed from the velocity of the tracer particles:
\begin{equation}
\sigma_{\rm v} \equiv  \left[{1 \over 3\,N}\,\sum_ {i=1}^{3} \, \sum_{n=1}^{N} ( u_{i,n} - \bar{u}_i )^2  \right ]^{1/2},
\label{eq_sigv}
\end{equation}
where $\bar{u}_i \equiv  \sum_{n=1}^{N} u_{i,n} /N$ are the components of the mean tracer particle velocity, and $N$ is the
total number of tracer particles in the cloud. We choose these definitions of radius and velocity dispersion because they relate directly to the
observable ones, so they yield estimates of the virial parameter that should be comparable to the observational values. Furthermore, we 
have previously found that the virial parameter based on these definitions of radius and velocity dispersion is very close to the actual energy 
ratio in realistic MCs, selected with our previous SN-driven simulation (see Sections 7 and 10.5 and equation (26) in Paper I).
Thus, in this work we use the effective virial parameter,
\begin{equation}
\alpha_{\rm vir,e} \equiv 5 \sigma^2_{\rm v} R_{\rm e}/(G M_{\rm cl}), 
\label{eq_alpha_e}
\end{equation}
and the effective dynamical time,
\begin{equation}
t_{\rm dyn,e} \equiv R_{\rm e}/\sigma_{\rm v,3D}.
\label{eq_tdyn_e}
\end{equation}
It is important to realize that $R_{\rm e}$ defined with the projected area is the same as that defined with the cloud volume, $V_{\rm cl}$,
$R_{\rm e,V}\equiv (V_{\rm cl}/(4\,\pi /3))^{1/3}$, only in the case of a spherical cloud. In general, $R_{\rm e}>R_{\rm e,V}$, so one should 
avoid estimating the mean cloud density as $M_{\rm cl}/(4\,\pi R_{\rm e}^3/3)$, as that would underestimate the cloud mean density and 
overestimate the cloud free-fall time and $SFR_{\rm ff}$. In our cloud samples we find that $\langle R_{\rm e}/R_{\rm e,V} \rangle \approx 1.6$
(giving a factor of two correction for the free-fall time), so the relation between $t_{\rm ff} / t_{\rm dyn,e}$ and $\alpha_{\rm vir,e}$ becomes:
\begin{equation}
{t_{\rm ff} \over t_{\rm dyn,e}} \approx {\pi \over 4} \sqrt{{3\over 10}}\, \alpha_{\rm vir,e}^{1/2} \approx 0.43\, \alpha_{\rm vir,e}^{1/2},
\label{eq_coeff}
\end{equation}
which differs by a factor of two from the coefficient derived from equation (\ref{eq_alpha}). Although we don't know the correction factor in the
case of real MCs, it is likely the free-fall time (hence $SFR_{\rm ff}$) is slightly overestimated as the cloud mean density is usually computed
as $M_{\rm cl}/(4\,\pi R_{\rm e}^3/3)$.

The Mach number is computed as in equation (\ref{eq_mach}), where $c_{\rm s}$ is derived from the mass-weighted average temperature
(the average value of the temperature associated to the tracers) and $\sigma_{\rm v,3D}=\sigma_{\rm v}\sqrt{3}$, using equation (\ref{eq_sigv}), 
which should yield values comparable to observational estimates of MC Mach numbers. We compute $\beta$ as the average of the local value 
of $\beta$ given by the ratio of thermal and magnetic pressure associated to each tracer particles, with a weight $m_{\rm i}/\rho_{\rm i}$ for each 
tracer, where $m_{\rm i}$ and $\rho_{\rm i}$ are the mass of the tracer and the local gas density at the tracer position. The weight is proportional 
to the volume of the gas element represented by the tracer, so we obtain a volume-averaged $\beta$. Observed values of $\beta$ may be 
estimated from the dispersion of the polarization angle in MCs. Due to the drop in polarization fraction with increasing density 
\citep[e.g.][]{Padoan+2001pol,Pelkonen+07,PlanckX!X15}, the derived $\beta$ is probably closer to a volume average value than a mass-weighted one.

The MC properties of both samples are plotted in Figure \ref{mc_properties}, where empty circles represent the clouds with $n_{\rm H,min}=200$ 
cm$^{-3}$ and filled circles the denser ones with $n_{\rm H,min}=400$ cm$^{-3}$. The two upper panels show the mass-size and velocity-size relations.
Our MCs have equivalent radii in the range 2.6-23.5 pc, and column densities in the range 10-80 M$_{\odot}$pc$^{-2}$. Despite the difference in the 
average column density and size of the two cloud samples, the slopes of the relations are essentially the same, and are also consistent with the 
observations and with results from our previous SN-driven simulation (see Papers I and II). The two lower panels of 
Figure \ref{mc_properties} show $\alpha_{\rm vir}$ versus $M_{\rm cl}$ (left) and $\beta$ versus $\sigma_{\rm v}$ (right). The virial parameter
spans a wide range of values, between approximately 0.5 and 25, with the maximum value (and the scatter) decreasing with increasing 
$M_{\rm cl}$. The gas to magnetic pressure is in the range 0.09-1.3, with an average value $\langle\beta\rangle=0.37$, and decreases 
with increasing $\sigma_{\rm v}$, showing that the magnetic field in the clouds is amplified by the cloud turbulence, as the thermal pressure does 
not vary much from cloud to cloud. However, the rms Alfv\'{e}nic Mach number, defined as ${\cal M}_{\rm a} \equiv \sqrt{{\cal M}^2\beta/2}$ and averaged 
over all clouds, is $\approx 3.1$, so the MC turbulence is super-Alfv\'{e}nic \citep{Padoan+Nordlund99mhd,Padoan+04power,Lunttila+08,Lunttila+09}. 

Another important non-dimensional MC parameter is the ratio, $\chi$, of the power in compressive and solenoidal modes of the velocity field, ${\bs v}$,
\begin{equation}
\chi \equiv \langle {\bs v}_{\rm c}^2 \rangle/  \langle {\bs v}_{\rm s}^2 \rangle, 
\label{eq_chi}
\end{equation}
where ${\bs v}_{\rm c}$ and ${\bs v}_{\rm s}$ are the compressive and solenoidal components of the velocity field.
The two velocity components are derived with the standard Helmholtz decomposition in Fourier space,
within a bounding box containing each MC, following the procedure in Paper II. We then compute the
turbulent compressive ratio, $\chi_{\rm t}$, by subtracting the mean cloud rotation, $V_{\rm r}$, and expansion, $V_{\rm e}$,
velocities, 
\begin{equation}
\chi_{\rm t} \equiv [\langle {\bs v}_{\rm c}^2 \rangle- V_{\rm e}^2]/[\langle {\bs v}_{\rm s}^2 \rangle - V_{\rm r}^2],
\label{eq_chit}
\end{equation}
where $V_{\rm r}$ and $V_{\rm e}$ are derived from the mean vorticity and the mean divergence in the cloud bounding 
box (see equations (1) and (2) in Paper II). As in the cloud samples from our previous SN-driven simulation,
we find lognormal distributions of $\chi_{\rm t}$, with both mean and standard deviations very close to our previous
values, $0.33\pm0.18$ for the clouds with $n_{\rm H,min}=200$ cm$^{-3}$, and $0.34\pm0.16$ for the clouds with 
$n_{\rm H,min}=400$ cm$^{-3}$. The non-dimensional MC parameters will be used in Section \ref{sect_pn11} to test the
predictions of the turbulent fragmentation model.

\section{Star Formation Rate in Molecular Clouds} \label{sec_sfr}

Each MC selected from the simulation is followed for 1.68 Myr after the time it is identified. This time is long enough to
evaluate both the time-averaged SFR and the time variations of the SFR of individual clouds, and short enough to avoid 
complications related to the identification of clouds as distinct objects in the turbulent flow. MCs are continuously 
being formed and dispersed by the SN-driven turbulence, and they are part of an interconnected filamentary structure
that extends up to the outer scale of the turbulence, approximately 70-100 pc \citep{Joung+09,deAvillez07scaling,Padoan+16SN_I}, 
so they can hardly be considered as well-defined and long-lived isolated entities. This complex ever-changing nature of MCs is not a 
major concern for this work, as our goal is to correlate the SFR with the physical parameter of MCs at a given time, rather than to 
follow their long-term evolution. 

\begin{figure}[t]
\includegraphics[width=\columnwidth]{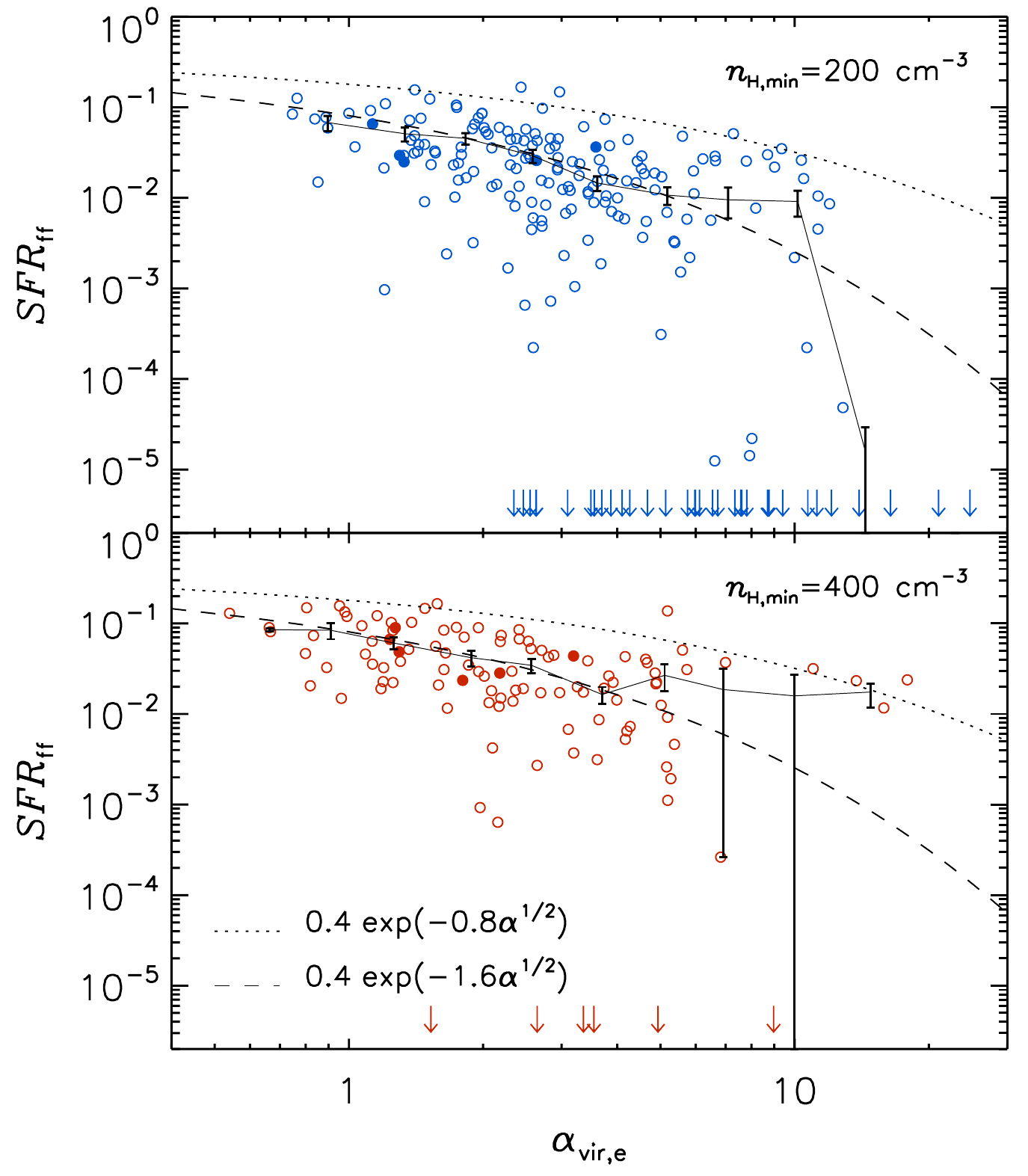}
\caption[]{SFR per free-fall time versus effective virial parameter for all the clouds selected from the simulation. The arrows show the values of $\alpha_{\rm vir,e}$
of the MCs with $SFR_{\rm ff}=0$. The error bars connected by the solid line are the values of $SFR_{\rm ff}$ averaged within logarithmic intervals of $\alpha_{\rm vir,e}$.
The dashed line is $SFR_{\rm ff,\alpha}$, the analytical fit to our revised PN11 model computed with the physical parameters of the MCs from the simulation, given by
equation (\ref{eq_sfr_ff_alpha}). The dotted line is a function like $\alpha_{\rm vir,\alpha}$, but with a smaller exponential coefficient, to trace roughly an upper envelope 
of the plot. The filled circles indicate the five MCs with $M_{\rm cl} > 2\times10^4$ M$_{\odot}$ (upper panel) and the six MCs with $M_{\rm cl} > 10^4$ M$_{\odot}$ 
(lower panel)}.
\label{sfr_alpha}
\end{figure}

To define the SFR over the time interval $\Delta t = 1.68$ Myr after the cloud identification, we track only the total mass 
in the tracer particles associated with the initial cloud definition. We don't account for the possibility that new gas is accreted 
to the cloud or removed from it during that time interval. The SFR is thus:
\begin{equation}
SFR \equiv  {  M_{\rm tr}(t_j) - M_{\rm tr}(t_j+\Delta t)   \over  \Delta t },
\label{eq_sfr}
\end{equation}
where, $M_{\rm tr}(t)$ is the total mass of tracers in the cloud that have not been accreted onto sink particles at the time $t$, and $t_j$ is 
the time corresponding to the $j$-th snapshot in which the cloud is identified. The SFR per free-fall time \citep{Krumholz+McKee05sfr} is then
defined as the following non-dimensional SFR:
\begin{equation}
SFR_{\rm ff} \equiv SFR \,/\, [ M_{\rm tr}(t_j) / t_{\rm ff} ],
\label{eq_sfr_ff}
\end{equation}
where $t_{\rm ff}$ is the free-fall time at the mean density of the cloud. To evaluate the fluctuations of $SFR_{\rm ff}$,
we also measure it within 14 shorter time intervals of length $\delta t = \Delta t / 14=0.12$ Myr.  

\begin{figure}[t]
\includegraphics[width=\columnwidth]{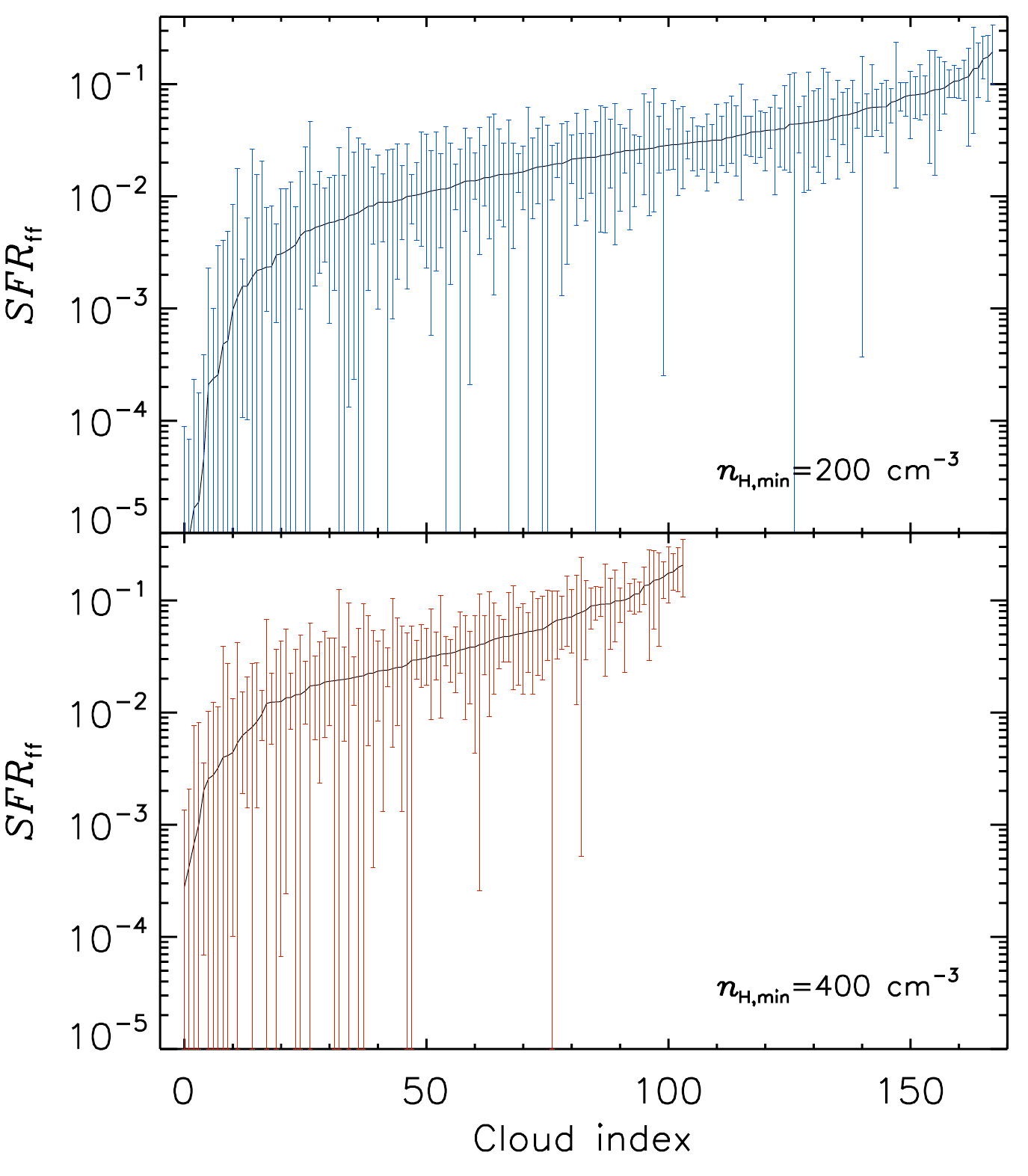}
\caption[]{Time variation of $SFR_{\rm ff}$ in each cloud, with clouds ordered by increasing value of their time-averaged $SFR_{\rm ff}$
(solid black line). Each error bar extends between the maximum and minimum SFR of a cloud, among the values measured within 14 time 
intervals of $\Delta t/14=0.12$ Myr size. Values of $10^{-5}$ indicate that the minimum SFR is zero. Clouds that never form stars during the
whole interval $\Delta t=1.68$ Myr are not shown in the figure (35 clouds in the catalog with $n_{\rm H,min}=200$ cm$^{-3}$ and 6 clouds
for $n_{\rm H,min}=400$ cm$^{-3}$.)}
\label{sfr_dispersion}
\end{figure}

Figure \ref{sfr_alpha} shows $SFR_{\rm ff}$ versus $\alpha_{\rm vir,e}$ for our two cloud samples (empty circles). There is a clear trend of decreasing 
SFR with increasing virial parameter, though with a very large scatter. The scatter increases with increasing $\alpha_{\rm vir,e}$ and must have primarily 
a random origin due to variations from cloud to cloud
at fixed virial parameter and time variations within each cloud, because we do not detect any strong dependence of $SFR_{\rm ff}$ on ${\cal M}$ or $\beta$ at 
constant $\alpha_{\rm vir,e}$ (and the scatter is much smaller in the case of the SFR predicted by our model using the cloud parameters, as shown in Figure 
\ref{sfr_alpha_model}). The origin of this scatter is discussed in Section \ref{sec_scatter}, where we present our model predictions. To illustrate the amount 
of time variations, we show in Figure \ref{sfr_dispersion} the maximum and minimum values of $SFR_{\rm ff}$ measured in the 14 time intervals $\delta t = 0.12$ 
Myr in each cloud. The characteristic scatter of $SFR_{\rm ff}$ due to time variations is over one order of magnitude.

The mean values of $SFR_{\rm ff}$ in
logarithmic bins of $\alpha_{\rm vir,e}$ are shown by the error-bar symbols in Figure \ref{sfr_alpha}, where the size of the error bars is equal to the 
standard error of the mean (the rms divided by square root of the the number of data points in the bin). Despite the sizable scatter, these mean values 
follow very closely the analytical fit to our model (see Section \ref{sec_model_pred}) applied to all the MCs selected from the simulation, 
$SFR_{\rm ff,\alpha} = 0.4 \, \exp(-1.6 \, \alpha_{\rm vir,e}^{1/2})$. This agreement is nearly equally good for both cloud samples, with the lower 
density one ($n_{\rm H,min}=200$ cm$^{-3}$) yielding only slightly smaller mean values (though always within approximately one standard error of 
the mean from the analytical fit of the model). The agreement with the model breaks down only at large values of $\alpha_{\rm vir,e}$,
where the statistical significance of the mean values of $SFR_{\rm ff}$ is low, due to the very small number of data points per bin
(the standard error of the mean can still be small, if most points are close to each other by chance).

\begin{figure}[t]
\includegraphics[width=\columnwidth]{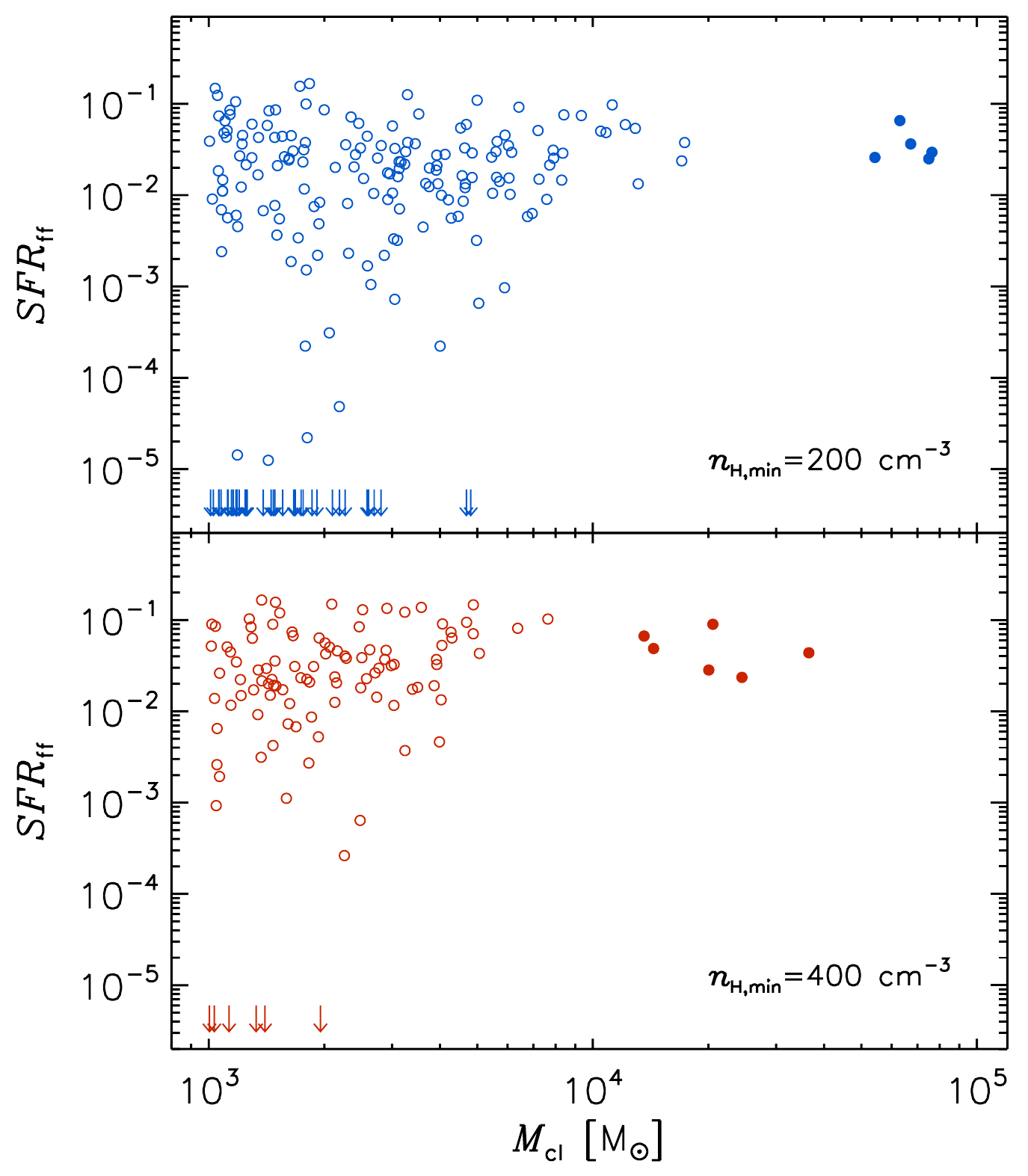}
\caption[]{SFR per free-fall versus cloud mass for the same MCs from the simulation as in the previous figures.}
\label{sfr_mass}
\end{figure}

The arrows in Figure \ref{sfr_alpha} indicate the values of $\alpha_{\rm vir,e}$ of the clouds where $SFR_{\rm ff}=0$. The upper panel shows that a significant 
fraction of clouds in the sample with $n_{\rm H,min}=200$ cm$^{-3}$ (approximately 18\%) have $SFR_{\rm ff}=0$, even at relatively low values of 
$\alpha_{\rm vir,e}$. The fraction drops to 5\% in the case of $n_{\rm H,min}=400$ cm$^{-3}$ (lower panel). We find that only relatively small clouds
have $SFR_{\rm ff}=0$. Figure \ref{sfr_mass} shows that $SFR_{\rm ff}=0$ only in clouds with $M_{\rm cl} < 5\times10^3$ M$_{\odot}$
or $M_{\rm cl} < 2\times10^3$ M$_{\odot}$ in the lower and higher $n_{\rm H,min}$ samples respectively. The lower envelope of the plots
in Figure \ref{sfr_mass} rises sharply with increasing cloud mass, so that the scatter in  $SFR_{\rm ff}$ is rather small for the most massive clouds.  
The average values for the 5 and 6 most massive clouds in the two samples (shown as filled circles in Figures \ref{sfr_alpha} and \ref{sfr_mass}) 
are $SFR_{\rm ff}=0.04\pm0.01$ ($n_{\rm H,min}=200$ cm$^{-3}$) and $SFR_{\rm ff}=0.05\pm0.02$ ($n_{\rm H,min}=400$ cm$^{-3}$). 

\begin{figure}[t]
\includegraphics[width=\columnwidth]{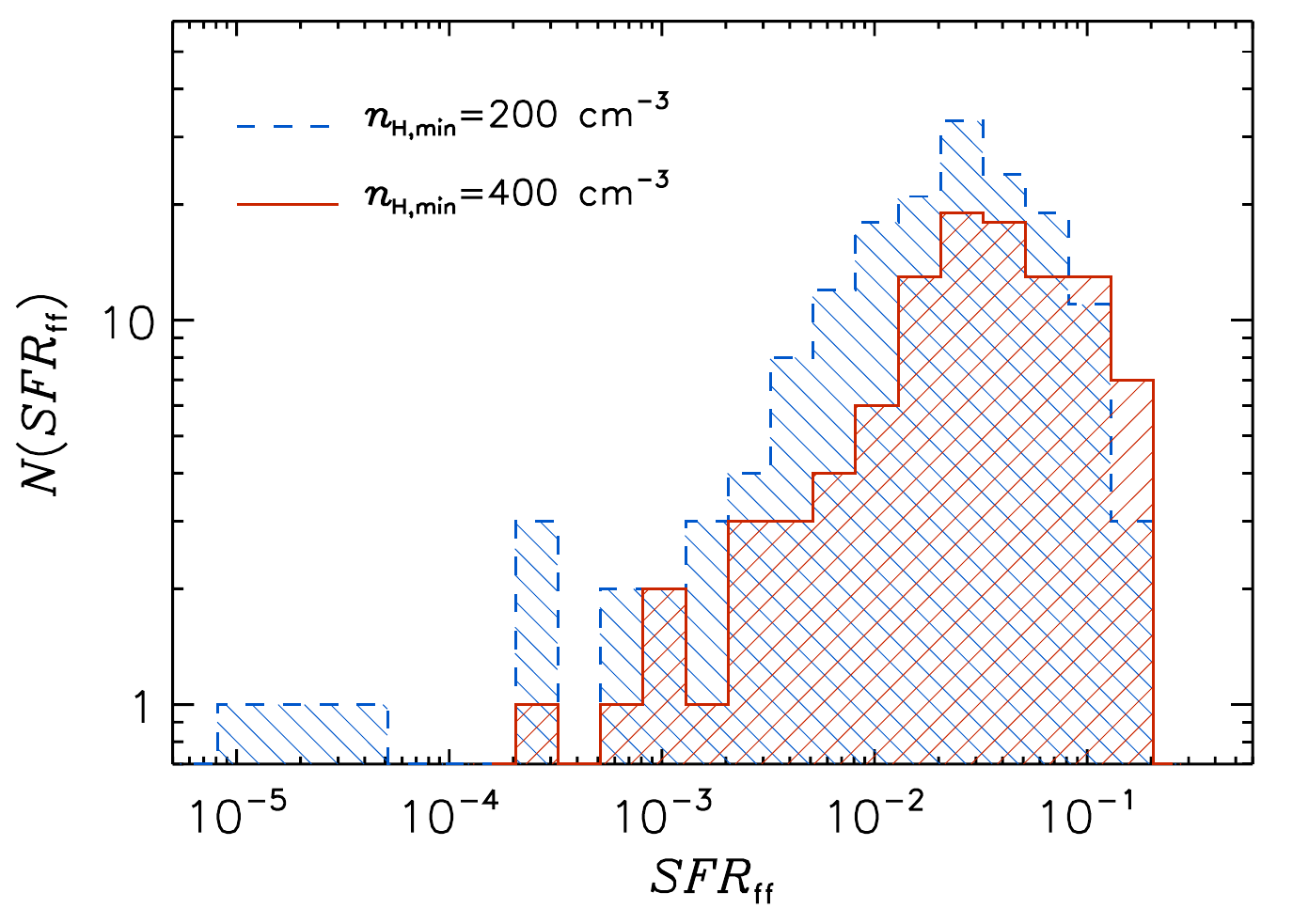}
\caption[]{Probability distributions of $SFR_{\rm ff}$ of our two samples of MCs extracted from the simulation.}
\label{sfr_distribution}
\end{figure}

The probability distributions of the SFR are shown in Figure \ref{sfr_distribution}. Although they both peak at $SFR_{\rm ff}\approx 0.025$, the denser clouds have, 
on the average, larger values, with $SFR_{\rm ff}=0.03\pm0.03$ for $n_{\rm H,min}=200$ cm$^{-3}$ and $SFR_{\rm ff}=0.04\pm0.04$ for $n_{\rm H,min}=400$ cm$^{-3}$.
These values are comparable to observational estimates of the SFR in MCs, as discussed below in Section \ref{sec_real_sfr}.

\begin{figure}[t]
\includegraphics[width=\columnwidth]{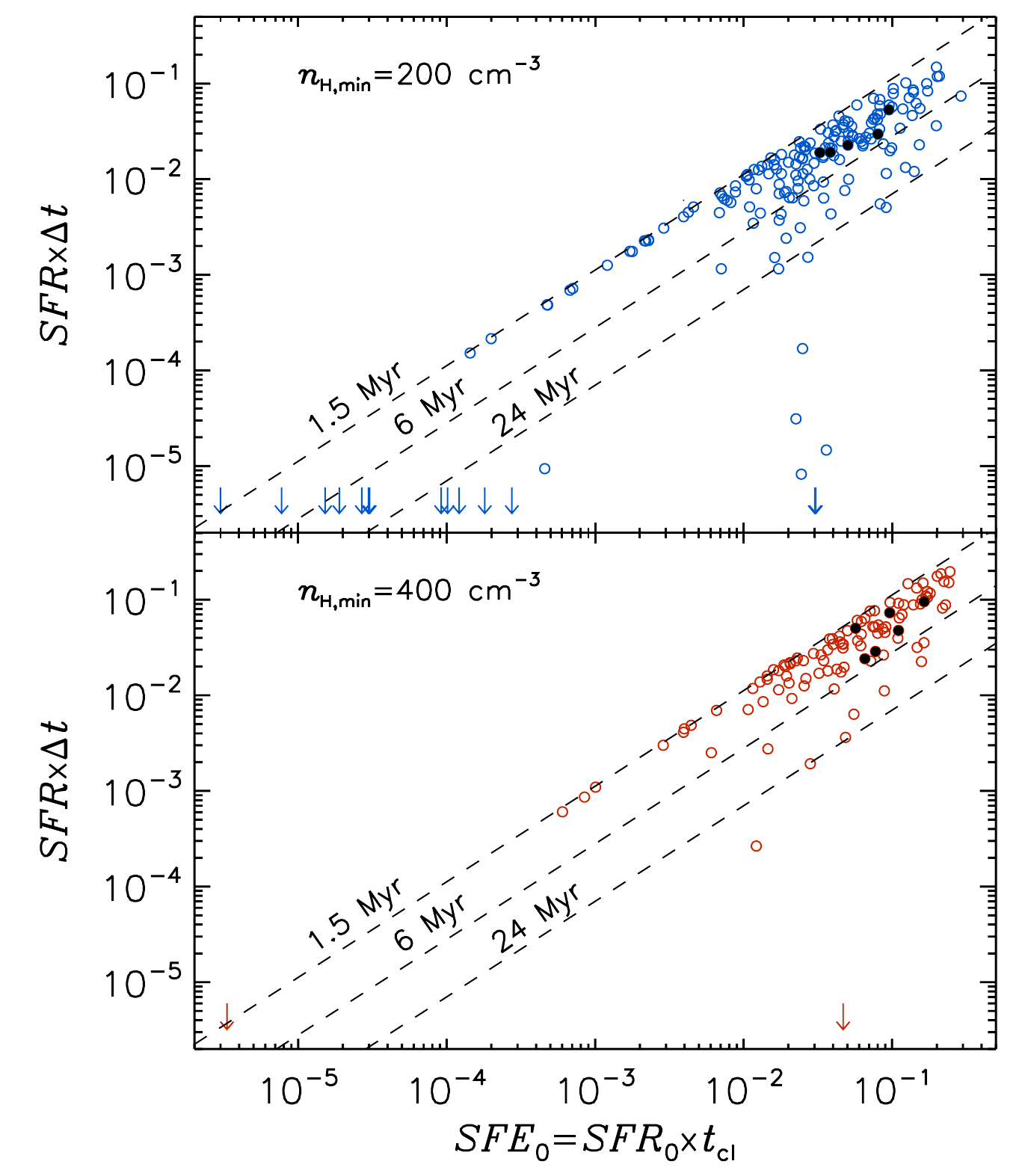}
\caption[]{Cloud mass fraction converted into stars in the time interval during which we follow the MCs, $\Delta t=1.68$ Myr, versus the cloud 
initial SFE, $SFE_0$. Assuming that the SFR prior to the time of the cloud selection, $SFR_0$, was constant and equal to the average one measured 
during $\Delta t$ after the cloud selection, $SFR_0=SFR$, we derive an estimate of the MC age at the time of selection, $t_{\rm cl}\equiv SFE_0/SFR$,
shown by the dashed lines for ages of 1.5, 6.0 and 24 Myr. As in previous figures, the filled circles are the five MCs with $M_{\rm cl} > 2\times10^4$ 
M$_{\odot}$ (upper panel) and the six MCs with $M_{\rm cl} > 10^4$ M$_{\odot}$ (lower panel). The arrows indicate the values of $SFE_0$ for
clouds with $SFR=0$.}
\label{sfr_sfe}
\end{figure}

In Figure \ref{sfr_sfe}, we plot the mass fraction turned into stars in the time interval $\Delta t=1.68$ Myr after the cloud is identified, which is given by 
$SFR\times \Delta t$, versus the initial star formation efficiency at the time the cloud is identified, $SFE_0$, given by
\begin{equation}
SFE_0 \equiv M_{\rm tr,s}(t_j)/(M_{\rm tr,s}(t_j)+M_{\rm tr}(t_j)),
\label{eq_sfe}
\end{equation}
where $M_{\rm tr,s}(t_j)$ is the total mass of the tracer particles inside the cloud and already accreted onto sinks at the time $t_j$ when the 
cloud is identified, and $M_{\rm tr}(t_j)$ is the total mass in tracers found in the same cloud and still in the gas phase. 

We can express the initial SFE as
$SFE_0=SFR_0\times t_{\rm cl}$, where $SFR_0$ is the average SFR up to the time the cloud is identified ($t<t_j$), and $t_{\rm cl}$ is the cloud age.
Although only some of the MCs have an approximately constant SFR during the time interval $\Delta t$, if we assumed that the average SFR prior to 
the cloud identification were equal to the average SFR after the cloud identification, $SFR_0=SFR$, we could derive the age of a cloud, $t_{\rm cl}$, 
at the time when it is selected, as $t_{\rm cl}\equiv SFE_0/SFR$. Constant values of $t_{\rm cl}$ are shown in Figure \ref{sfr_sfe} by dashed lines. 
The upper envelope of the plot in Figure \ref{sfr_sfe} is very well described by 
a line of constant cloud age, $t_{\rm cl} \approx 1.5$ Myr. Thus, if the SFR prior to the time of the cloud identification were constant and equal to the 
average SFR measured after the cloud identification, none of the cloud selected from the simulation would be younger than approximately 1.5 Myr,
while most clouds would be younger than 6 Myr, and almost every cloud younger than 24 Myr. As for the few clouds that would correspond to even 
larger age (including those with $SFR=0$ and $SFE_0>0$), it is more likely that the approximation of constant SFR does not apply and their SFR 
has declined with time, $SFR_0 >  SFR$, rather than their age being much larger than 20 Myr. 

The MC lifetime was studied in Paper I, using our previous SN-driven simulation, where we found that it was approximately four dynamical times
on average, $t_{\rm life}/t_{\rm dyn} = 4.5 \pm 1.7$, resulting in cloud lifetimes consistent with those estimated for MCs in the Large Magellanic 
Cloud \citep{Kawamura+09LMC}. Using the average cloud dynamical times in our two cloud samples, we would estimate an average cloud lifetime
$t_{\rm life}=14.6\pm4.5$ Myr for $n_{\rm H,min}=200$ cm$^{-3}$ and $t_{\rm life}=11.1\pm3.5$ Myr for $n_{\rm H,min}=200$ cm$^{-3}$. Most clouds
in the two samples have smaller estimated ages than such lifetimes, suggesting that most of the ages derived from the assumption of constant SFR are
reasonable.

\section{The Turbulent Fragmentation Model}\label{sect_pn11}

The SFR can be modeled as the result of the gas fragmentation by the supersonic turbulence 
\citep{Padoan95,Krumholz+McKee05sfr,Padoan+Nordlund11sfr,Federrath+Klessen12,Hopkins13}.
Given the probability density function (PDF) of the gas density in supersonic MHD turbulence
and a critical density for collapse, the integral of the PDF above the critical density, divided by
an appropriate time-scale, provides an estimate of the SFR. Different models differ from each other
in modeling the PDF, in the definition of the critical density and in the choice of the time-scale,
and were recently reviewed in \citet{Padoan+14PPVI}. In this work, we only consider the model
by \citet[][PN11 hereafter]{Padoan+Nordlund11sfr}, which we slightly revise, motivated both by physics 
considerations and by the numerical results. In the following subsections, we first summarize the PN11 
model and present our revision; we then apply the revised model to the clouds selected from the simulation
and discuss the origin of the scatter in the $SFR_{\rm ff}$--$\alpha_{\rm vir}$ relation.

\subsection{The PN11 Star Formation Rate Model}

The model depends on the three non-dimensional parameters, $\alpha_{\rm vir}$, ${\cal M}$ and $\beta$, 
that express the ratios of turbulent kinetic, gravitational, thermal and magnetic energies. These parameters were 
defined and computed in Section \ref{sect_mcs} for all the clouds of our two samples. In the model revision described 
below, we introduce a fourth non-dimensional parameter, $\chi_{\rm t}$, also defined in Section \ref{sect_mcs}.

The critical density is defined as the external density of a critical Bonnor-Ebert sphere of size equal to the characteristic thickness of the
postshock gas in the turbulent flow. Using the MHD jump conditions of an isothermal shock, PN11 derive the following expression:
\begin{equation}
     {\rho_{\rm cr}\over \rho_0} = 0.067 \, \theta^{-2} \alpha_{\rm vir} \, {\cal M}^2
     {(1+0.925\beta^{-{3\over 2}})^{2\over 3} \over (1+\beta^{-1})^2},
\label{eq_rhocrmhd}
\end{equation}
where $\rho_0$ is the mean gas density, and $\theta$ is the fraction of the cloud diameter corresponding to the characteristic size of the 
largest compressive motions in the turbulent flow, with $\theta=0.35$ in PN11.

The density PDF is assumed to be a lognormal distribution,
\begin{equation}
p(x)dx=\frac{ 1 } { x\,(2\pi\sigma^2)^{1/2}}\, {\rm exp} \left[ -\frac{({\rm ln}x + \sigma^2/2)^2}{2\, \sigma^2} \right] dx,
\label{eq_pdf_mhd}
\vspace{0.2cm}
\end{equation}
where $x=\rho/\rho_0$ is the gas density normalized to the mean and $\sigma$ is the standard deviation of ln$(x)$ that depends on both ${\cal M}$ and $\beta$, 
\begin{equation}
\sigma^2\approx {\rm ln}\left[  1 +  \left(b\,{\cal M}\right)^2 (1+\beta^{-1})^{-1} \right],
\label{eq_sigma_mhd}
\end{equation}
corresponding to the following standard deviation for $x$,
\begin{equation}
\sigma_{x}\approx (1+\beta^{-1})^{-1/2} b\,{\cal M}.
\label{eq_sigma_x}
\end{equation}
In PN11, $b=0.5$ and $\beta=0.39$, based on numerical estimates and fitting of the PDF. In Section \ref{sect_mcs}, we have shown that
the average value of beta for the clouds selected from the simulation is $\langle \beta \rangle =0.39$, while the average value of $b$, 
using its dependence on $\chi_{\rm t}$ given below and proposed in Paper II, and the distribution of $\chi_{\rm t}$ for our clouds, is
$\langle b \rangle =0.48$. Thus, the values of $b$ and $\beta$ adopted in PN11 were very close to the ones derived in this work
for a realistic sample of MCs.

Assuming that a fraction $\epsilon$ of the mass fraction above the critical density is turned into stars in a free-fall time of the
critical density, $t_{\rm ff,cr}=(3 \pi/(32 G \rho_{\rm cr}))^{1/2}$, the star formation rate per free-fall time (the
mass fraction turned into stars in a free-fall time) is given by:
\begin{eqnarray}
SFR_{\rm ff,PN11} & = & \epsilon \, {t_{\rm ff}\over t_{\rm ff,cr}} \int_{x_{\rm cr}}^\infty x \,p(x)\, dx  \nonumber \\
                                & = & \epsilon \, \frac{x_{\rm cr}^{1/2}}{2}\left(1+{\rm erf} \left[  \frac{\sigma^2-2\,{\rm ln}\left(x_{\rm cr}\right) }
                                {2^{3/2}\,\sigma}\right]\right)
\label{eq_pn11}
\end{eqnarray}
where $t_{\rm ff}=(3 \pi/(32 G \rho_0))^{1/2}$ is the free-fall time of the mean density and $x_{\rm cr}=\rho_{\rm cr}/\rho_0$.

\subsection{Revision of the PN11 SFR Model}  \label{sect_rev}

\begin{figure}[t]
\includegraphics[width=\columnwidth]{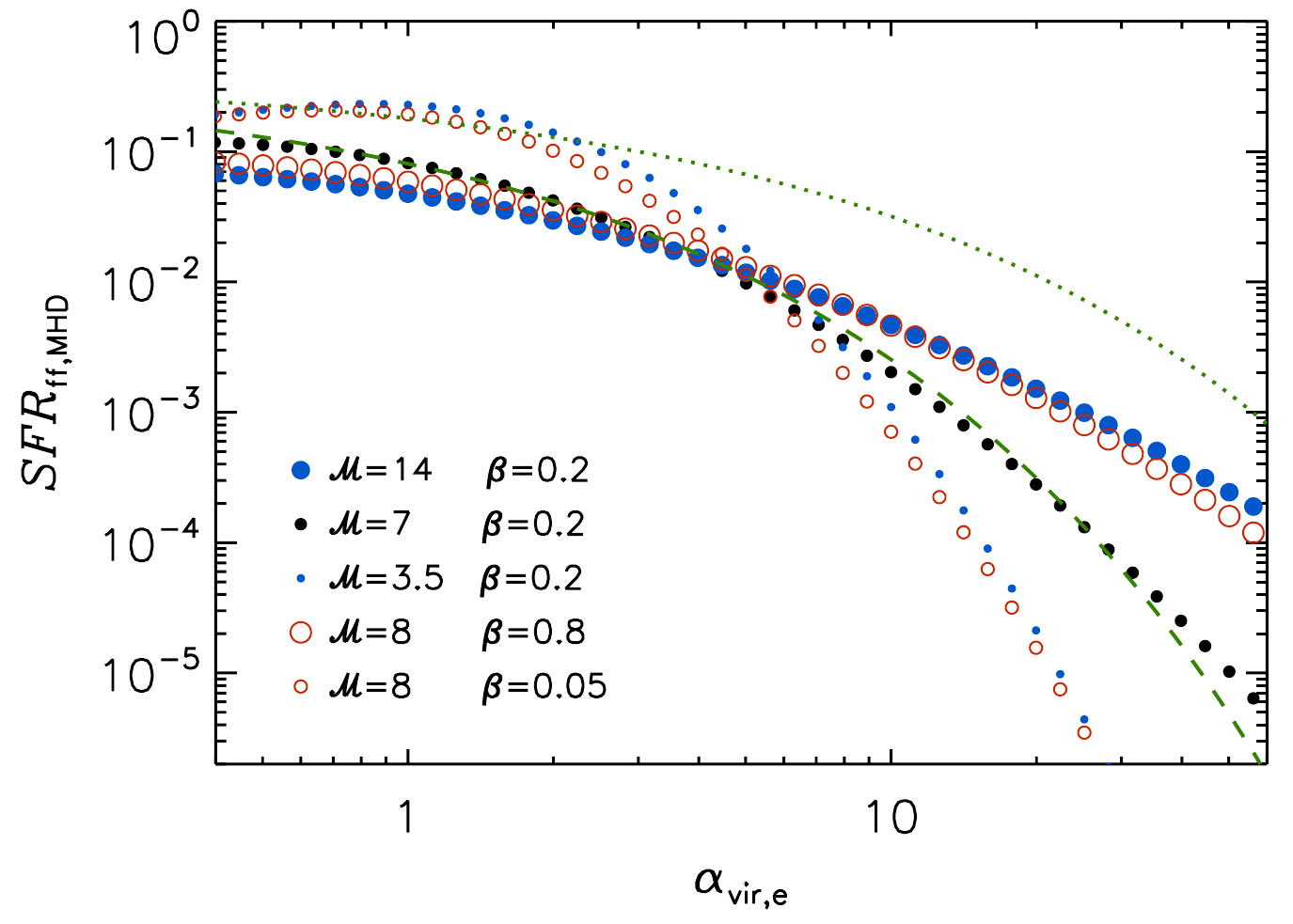}
\caption[]{SFR per free-fall time versus virial parameter predicted by our revision of the PN11 model, for five combinations of four values 
of ${\cal M}$ and three values of $\beta$, assuming $\epsilon=0.5$ and $b=0.48$ (the latter is the average value from our cloud samples). 
The dashed line shows the analytical fit to our model applied to the MCs from the simulation, $SFR_{\rm ff,\alpha}$ (see Section \ref{sec_model_pred} 
and Figure \ref{sfr_alpha_model}), and the dotted line the same function, but with a smaller exponential coefficient, that was used in Figure \ref{sfr_alpha} 
to trace an approximate upper limit of the $SFR_{\rm ff}$-$\alpha_{\rm vir,e}$ relation.}
\label{sfr_model}
\end{figure}

The first modification to our model is the choice of the time-scale that defines the SFR. While in PN11 we assumed that the timescale
was the free-fall time of the critical density, $t_{\rm ff,cr}$, here we choose the time of formation of the high-density
tail of the PDF, $t_{\rm PDF}$, which we define as the lifetime of compressions responsible for the characteristic postshock density 
used in our derivation of the critical density, and we also assume $R_{\rm cl}=R_{\rm e}$,
\begin{equation}
t_{\rm PDF} \equiv 2\,\theta \, R_{\rm e} / \sigma_{\rm v,3D} = 2\,\theta \, t_{\rm dyn,e}.  
\label{eq_tpdf}
\end{equation}
Thus, the coefficient of $SFR_{\rm ff}$ is $t_{\rm ff} / t_{\rm PDF}$ instead of $t_{\rm ff} / t_{\rm ff,cr}$.
The free-fall time of the critical density is generally too short to assume that the high-density tail of the
PDF is maintained despite the collapse of the dense gas. In our cloud samples, 
$\langle t_{\rm PDF}/ t_{\rm ff,cr} \rangle \approx 3.8$, so the average SFR is set by $t_{\rm PDF}$
rather than $t_{\rm ff,cr}$, while $t_{\rm ff,cr}$ (and the unrevised PN11 model) sets the maximum value 
of the SFR, when the PDF tail is fully sampled.

The ratio $t_{\rm ff} / t_{\rm PDF}$ can be related to $\alpha_{\rm vir,e}$ using equation (\ref{eq_coeff}):
\begin{equation}
{t_{\rm ff} \over t_{\rm PDF}} \approx {\pi \over 8\, \theta} \sqrt{{3\over 10}}\, \alpha_{\rm vir,e}^{1/2} \approx 0.215\, \theta^{-1} \, \alpha_{\rm vir,e}^{1/2},
\label{eq_coeff2}
\end{equation}
so our revised expression for the SFR per free-fall time is:
\begin{equation}
SFR_{\rm ff,MHD} = 0.215 \,\epsilon \,\theta^{-1} \alpha_{\rm vir,e}^{1/2}\left(1+{\rm erf} \left[  \frac{\sigma^2-2\,{\rm ln}\left(x_{\rm cr}\right) }
                                {2^{3/2}\,\sigma}\right]\right).
\label{eq_sfr_ff}
\end{equation}

The second modification is in the expression for the critical density. Because it is based on the characteristic postshock density,
it makes sense to consider the product $b\, {\cal M}$ instead of ${\cal M}/2$, in analogy with the derivation of the standard deviation
of the density PDF. Furthermore, because we can measure the ratio, $\chi_{\rm t}$, of the power in compressive and solenoidal modes 
for each cloud, we relate the parameter $b$ to $\chi_{\rm t}$ by assuming, as in Paper II, that $b\, {\cal M}$ is the rms Mach number 
of the compressive part of the velocity field, which gives
\begin{equation}
b = \sqrt{\chi_{\rm t} / (1+\chi_{\rm t})}.
\label{eq_bpar}
\end{equation}
With this final modification, $b$ and ${\cal M}$ only appear in their product, so a change in $b$ is equivalent to a change in ${\cal M}$.

The revised version of the critical density is:
\begin{equation}
     x_{\rm cr} = 0.268 \, \theta^{-2} \alpha_{\rm vir,e} \,{\chi_{\rm t} \over (1+\chi_{\rm t})} \, {\cal M}^2
     {(1+0.925\beta^{-{3\over 2}})^{2\over 3} \over (1+\beta^{-1})^2},
\label{eq_rhocrmhd2}
\end{equation}

This revised model is illustrated in Figure \ref{sfr_model}, for five combinations of four values of ${\cal M}$ and three values of $\beta$, 
assuming $\epsilon=0.5$ and $b=0.48$, the average value of our cloud sample. The dashed line shows the analytical fit to the model applied 
to the MCs from the simulation, $SFR_{\rm ff,\alpha} = 0.4 \, \exp(-1.6 \, \alpha_{\rm vir,e}^{1/2})$ (see Section \ref{sec_model_pred}), as in 
Figure \ref{sfr_alpha}. The reference model (medium-size filled circles) with the parameter values corresponding to the peak (not the average) 
of the probability distributions of ${\cal M}$ and $\beta$, ${\cal M}=7$ and $\beta=0.2$, is nearly indistinguishable from the analytical fit, 
$SFR_{\rm ff,\alpha}$. The figure also shows that a variation of ${\cal M}$ by a factor $f$ corresponds approximately to a variation of 
$\beta$ by a factor $f^2$. This can be easily seen in the limit of $\beta \ll 1$, where both coefficients containing $\beta$ in equations 
(\ref{eq_rhocrmhd}) and (\ref{eq_sigma_mhd}) become $\approx \beta$, so $\beta$ and ${\cal M}$ only appear in the product $\beta\,{\cal M}^2$, 
which explains the observed dependence on these parameters in Figure \ref{sfr_model}. Thus, in the limit of $\beta \ll1$, our revised 
model depends only on two parameters, the virial parameter, $\alpha_{\rm vir,e}$, and the effective Mach number, ${\cal M}_{\rm e}$,
\begin{equation}
{\cal M}_{\rm e} \equiv b\,{\cal M}\,\beta^{1/2}=(2/\gamma)^{1/2}b\,{\cal M}_{\rm A}
\label{eq_mach_e}
\end{equation}
The second equality, derived from the definition of $\beta$ in eq. (\ref{eq_beta}), shows that, in the limit $\beta \ll1$, the effective Mach
number is the compressive part of the Alfv{\'e}nic Mach number.
Furthermore, at a value of $\alpha_{\rm vir,e}\approx 5-6$, $SFR_{\rm ff,MHD}$ is nearly independent of ${\cal M}_{\rm e}$, as shown in 
Figure \ref{sfr_model}, while it increases with increasing ${\cal M}_{\rm e}$ for $\alpha_{\rm vir,e} \gtrsim 6$ and with decreasing 
${\cal M}_{\rm e}$ for $\alpha_{\rm vir,e} \lesssim 5$.

The PN11 model was originally derived under the assumption of an isothermal equation of state, while the current simulation,
as well as the real ISM, is not isothermal. However, most of the mass of a MC is at gas densities such that the characteristic dynamical 
times are shorter than the gas cooling time, which justifies the isothermal approximation. Although a significant fraction of the volume 
within the outer boundaries of a MC may contain warm, low-density gas, the total momentum of that warm gas is negligible, thus it 
is not expected to affect significantly the high-density tail of the density PDF nor the SFR of the cloud.

\subsection{Model Predictions for the Numerical MC Samples} \label{sec_model_pred}

\begin{figure}[t]
\includegraphics[width=\columnwidth]{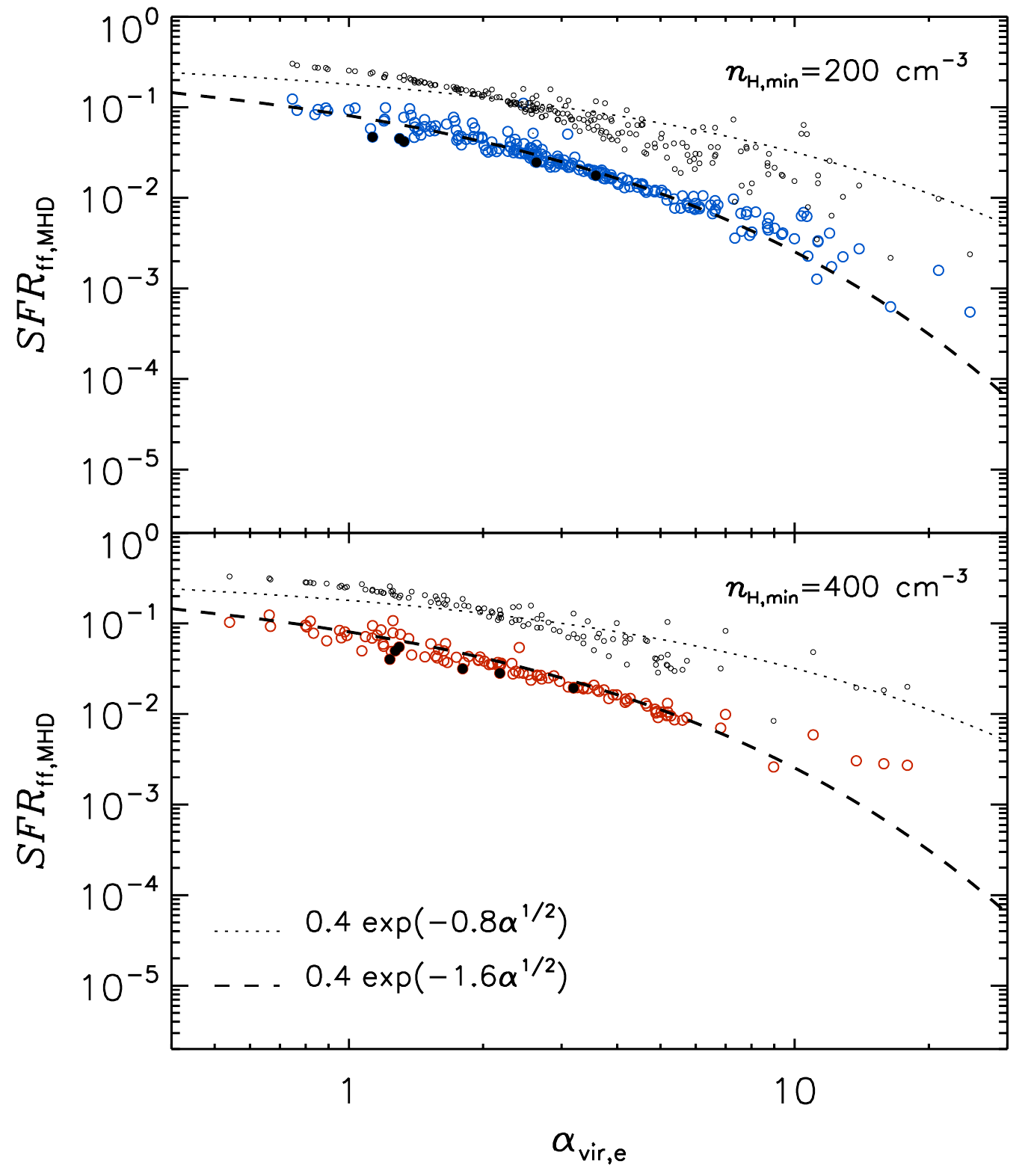}
\caption[]{SFR per free-fall time versus effective virial parameter predicted by our model (equation (\ref{eq_sfr_ff})) for the physical parameters 
of the MCs extracted from the simulation (large empty circles). The filled circles show the five (upper panel) and six (lower panel) most massive 
MCs, as in previous figures. The dashed line is the fit to the clouds with $\alpha_{\rm vir,e} < 10$, $SFR_{\rm ff,\alpha}$, given in equation
(\ref{eq_sfr_ff_alpha}). The dotted line is the same function, but with a smaller exponential coefficient, that was used in Figure \ref{sfr_alpha} 
to trace an approximate upper limit of the $SFR_{\rm ff}$-$\alpha_{\rm vir,e}$ relation. The small empty circles are upper limits for each cloud 
predicted by the model (see text in Section \ref{sec_model_pred}).}
\label{sfr_alpha_model}
\end{figure}

\begin{figure}[t]
\includegraphics[width=\columnwidth]{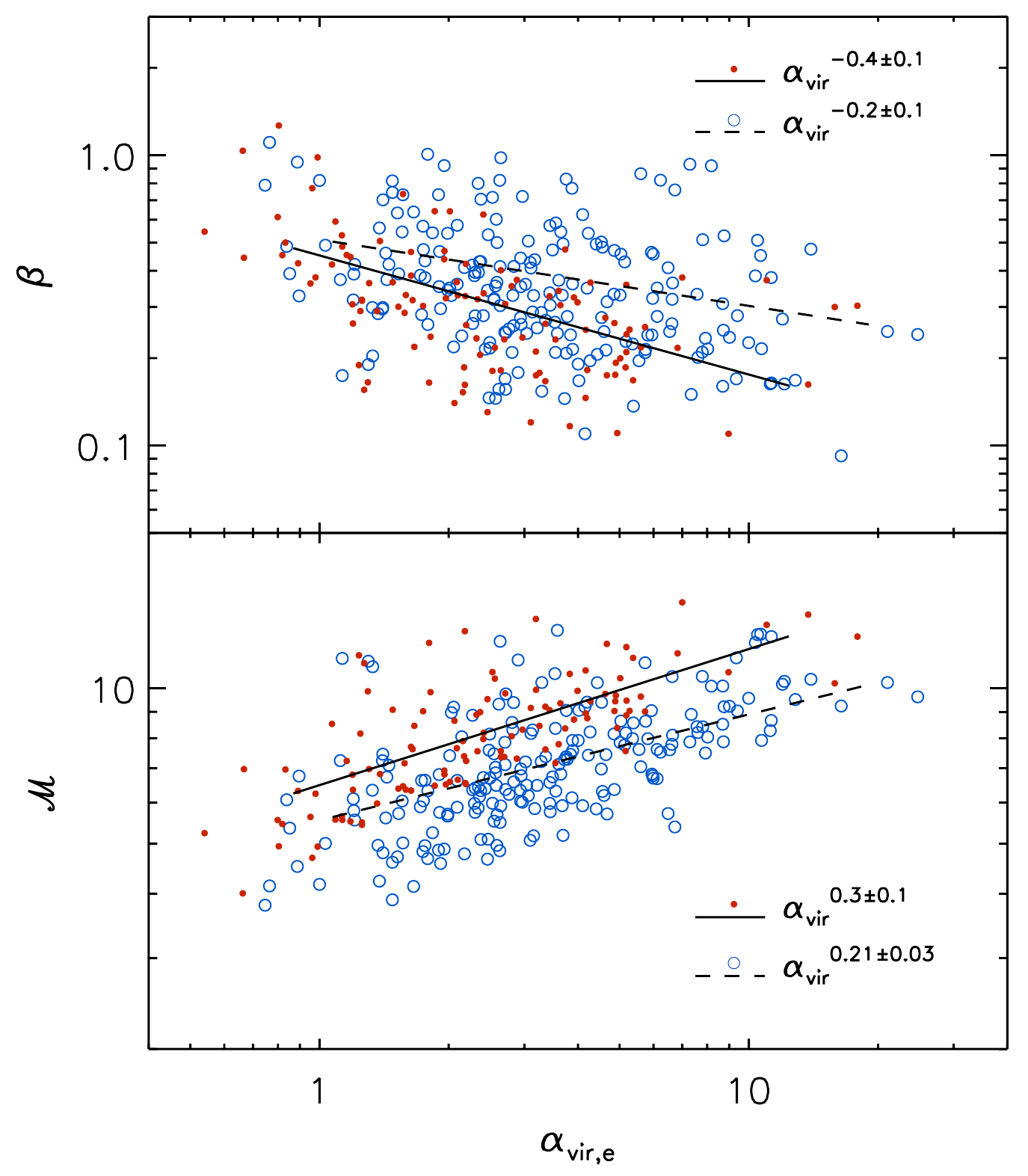}
\caption[]{Ratio of gas pressure to magnetic pressure (upper panel) and rms Mach number (lower panel) versus effective virial parameter for the MCs
selected in the simulation. The solid and dashed lines are the power-law fits to the mean values of $\beta$ and ${\cal M}$ averaged inside logarithmic 
intervals of $\alpha_{\rm vir,e}$, for the cloud samples with $n_{\rm H,min}=200$ (solid lines) and 400 (dashed lines) cm$^{-3}$.}
\label{beta_Mach_versus_alpha}
\end{figure}

We apply the revised PN11 model to the physical parameters of the numerical samples of MCs derived in Section \ref{sect_mcs}. 
The dependence of the predicted SFR, $SFR_{\rm ff,MHD}$, on $\alpha_{\rm vir,e}$ is shown in Figure \ref{sfr_alpha_model} 
(large empty circles). Despite the wide range of parameter values, the predicted $SFR_{\rm ff,MHD}$-$\alpha_{\rm vir,e}$
relation has a very small scatter, compared with that of the corresponding numerical relation in Figure \ref{sfr_alpha}.
This is partly due to the anti-correlation between ${\cal M}$ and $\beta$, as shown in Figure \ref{beta_Mach_versus_alpha},
resulting in a small scatter in ${\cal M}_{\rm e}$. The average and standard deviation of ${\cal M}_{\rm e}$ are $2.1\pm0.7$ and $2.2\pm0.7$ 
in the lower and higher $n_{\rm H,min}$ samples respectively, which explains the small scatter in Figure \ref{sfr_alpha_model}.

Because the predicted $SFR_{\rm ff,MHD}$-$\alpha_{\rm vir,e}$ relation does not show a significant dependence on $n_{\rm H,min}$, 
and given its small scatter, we provide an analytical fit inspired by our earlier results from a large set of idealized turbulence simulations
\citep{Padoan+12sfr}: 
\begin{equation}
SFR_{\rm ff,\alpha} = 0.4 \, \exp(-1.6 \, \alpha_{\rm vir,e}^{1/2}),
\label{eq_sfr_ff_alpha}
\end{equation}
shown by the dashed line in Figure \ref{sfr_alpha_model}. In Section \ref{sec_sfr} we showed that $SFR_{\rm ff,\alpha}$ is also an excellent
fit to the values of $SFR_{\rm ff}$ from the simulation averaged within logarithmic intervals of $\alpha_{\rm vir,e}$ (Figure  \ref{sfr_alpha}). 
Thus, the model predicts successfully the average SFR, but without describing its scatter. We also showed, in Section \ref{sect_rev}, that
 $SFR_{\rm ff,\alpha}$ is nearly indistinguishable from the revised model, $SFR_{\rm ff,MHD}$, with a fixed set of parameters, $\epsilon=0.5$,
 $b=0.48$, $\beta=0.2$ and ${\cal M}=7$ (see Figure \ref{sfr_model}). As shown in Figure \ref{beta_Mach_versus_alpha}, 
${\cal M}$ and $\beta$ are correlated with $\alpha_{\rm vir,e}$, in such a way that the product ${\cal M}\,\beta^{1/2}$
is nearly independent of $\alpha_{\rm vir,e}$. Furthermore, we find that $b$ is also independent of $\alpha_{\rm vir,e}$, so
the effective Mach number, ${\cal M}_{\rm e}$, is also approximately independent of $\alpha_{\rm vir,e}$, and variations of ${\cal M}_{\rm e}$
from cloud to cloud cannot modify the relation $SFR_{\rm ff,MHD}$--$\alpha_{\rm vir,e}$, but only contribute to its (small) scatter. 
This is the reason why a single set of parameter values provides a relatively good fit to the average SFR of all clouds.


In \citet{Padoan+12sfr}, using a large set of AMR simulations, we derived a SFR law that depended only on the ratio 
$t_{\rm ff} / t_{\rm dyn}$ or, equivalently, on $\alpha_{\rm vir}$, $SFR_{\rm ff,P12} = \epsilon \, \exp(-1.6 \, t_{\rm ff} / t_{\rm dyn})$. 
Using equation (\ref{eq_alpha}) and assuming that the efficiency factor is $\epsilon=0.5$, we would obtain 
$SFR_{\rm ff,P12} = 0.5 \, \exp(-1.38 \, \alpha_{\rm vir}^{1/2})$, so the coefficients would be different from those of the
analytical fit in this work, $SFR_{\rm ff,\alpha}$. However, the value of $t_{\rm ff} / t_{\rm dyn}$ in \citet{Padoan+12sfr} 
was the average over the whole computational volume, while most of the star formation in those simulations occurred in dense clumps where 
the local value of $t_{\rm ff} / t_{\rm dyn}$ (or $\alpha_{\rm vir}$) may have been smaller than the mean value. The results in 
\citet{Padoan+12sfr} are not necessarily inconsistent with the current results\footnote{They would be consistent if, for example, the star-forming clumps
in the turbulence simulations had a virial parameter on average approximately 30\% smaller than the global one.}, but relating the SFR in those 
idealized turbulence simulations to the SFR in real MCs is an open problem. The need for a more realistic numerical sample of star-forming clouds 
was one of the motivations for the current work. On the other hand, results from a general cubic region of the ISM, as in \citet{Padoan+12sfr}, are
more relevant if the goal is to develop a subgrid model for star formation for galaxy formation simulations \cite[e.g.][]{Semenov+16}.

\begin{figure}[t]
\includegraphics[width=\columnwidth]{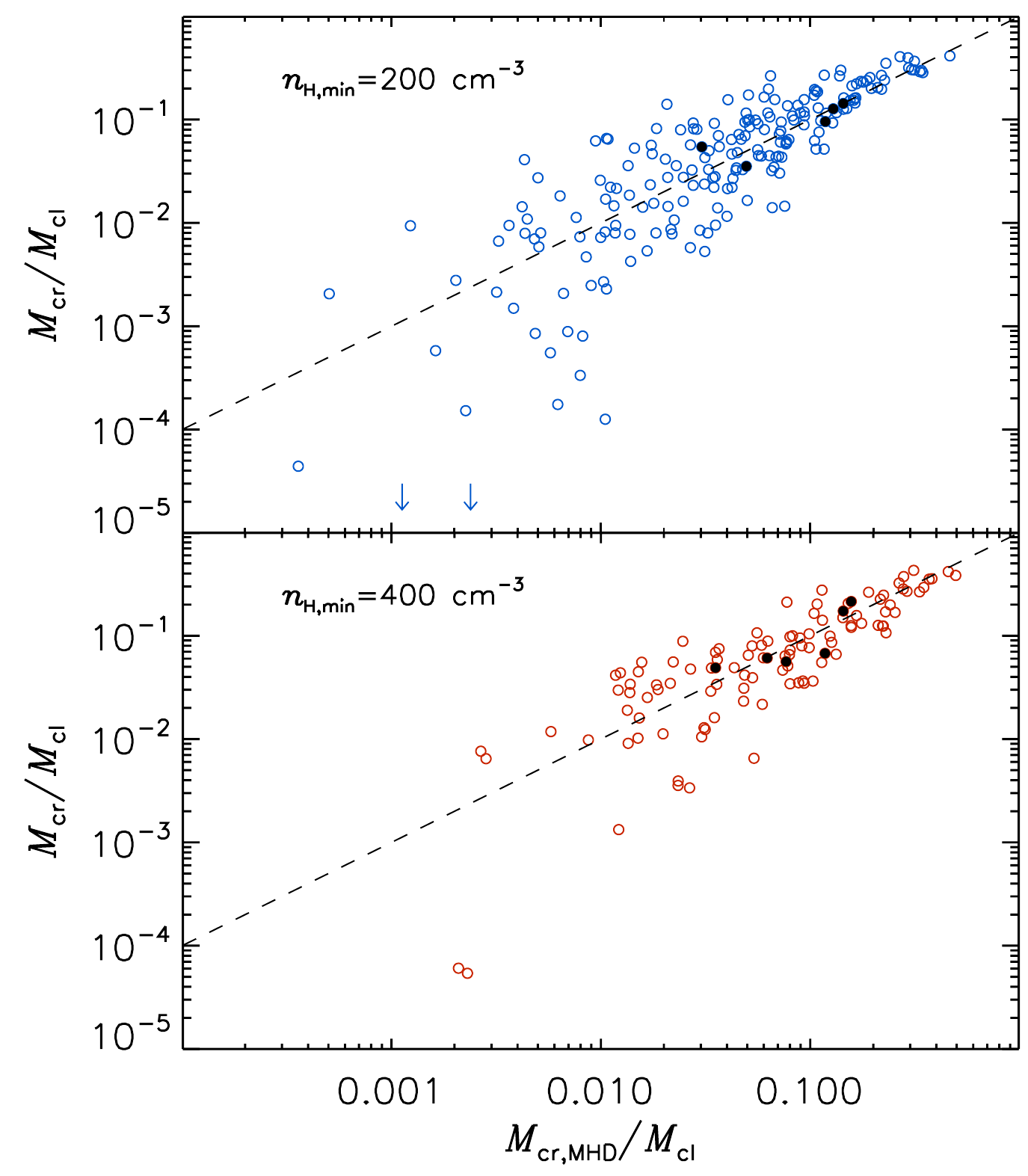}
\caption[]{Measured versus predicted mass fraction above the critical density for every cloud in the two numerical catalogs. $M_{\rm cr}$ 
is the cloud mass above the critical density measured in the simulation at the time of cloud selection; $M_{\rm cr,MHD}$ is the cloud mass 
above the critical density predicted by the model.}
\label{sfr_pdf_sfr_mhd}
\end{figure}

\subsection{The Scatter in the $SFR_{\rm ff}$--$\alpha_{\rm vir,e}$ Relation}\label{sec_scatter}

The comparison of Figures \ref{sfr_alpha} and \ref{sfr_alpha_model} shows that the scatter in the values of $SFR_{\rm ff}$ at constant
$\alpha_{\rm vir,e}$ from the simulation is much larger than the predicted scatter from our revised PN11 model, so the variations in
$b$, $\beta$ or ${\cal M}$ from cloud to cloud can explain only a small fraction of the scatter. However, large variations from cloud to cloud
and time variations in individual clouds are not inconsistent with the physical assumptions of the model. The average ratio between the formation time of dense structures
in the revised model and their collapse time is $\langle t_{\rm PDF}/t_{\rm ff,cr} \rangle\approx 3.8$, where the value is computed over both
cloud samples. Thus, there could be periods of high SFR, lasting for a time of order $\langle t_{\rm ff,cr} \rangle\approx 0.6$ Myr (the average is again 
over both cloud samples), followed by periods with very low (or zero) SFR, lasting for a time of order $\langle t_{\rm PDF} \rangle\approx 2$ Myr,
during which the depleted high density tail of the PDF grows back to its expected level. This intermittent behavior is more likely to occur in regions where
the SFR is very low, such as in small clouds and for high values of $\alpha_{\rm vir,e}$ (which are also more likely to occur in lower mass clouds),
because of the small number of collapsing cores. In very large clouds and at small values of $\alpha_{\rm vir,e}$, the number of cores is very large
and time variations of the high density tail of the PDF should have a smaller amplitude. In Figure \ref{sfr_alpha}, the scatter indeed
decreases towards smaller values of $\alpha_{\rm vir,e}$.

The model is consistent with finding clouds with no star formation, particularly at large $\alpha_{\rm vir,e}$, if the PDF tail is completely depleted.
The maximum SFR can be estimated from the integral of the PDF above the critical density, assuming the PDF tail is fully sampled, divided by 
the collapse time, so the coefficient of $SFR_{\rm ff}$ is $t_{\rm ff}/t_{\rm ff,cr}$, as in PN11, instead of $t_{\rm ff}/t_{\rm PDF}$
as in the revised model. The estimated values of the maximum SFR for all the clouds in our samples are shown in Figure \ref{sfr_alpha_model} 
(small empty circles). They are almost a factor of three to four larger than the predicted mean values (very close to the prediction of the PN11 
model) and follow approximately the dotted line that was shown to be an approximate upper envelope of the $SFR_{\rm ff}$-$\alpha_{\rm vir,e}$ 
relation in Figure \ref{sfr_alpha}. At $\alpha_{\rm vir,e}<1$, the maximum values are systematically above the dotted line, while the $SFR_{\rm ff}$
values from the simulation are all below the line, consistent with our expectation that the scatter caused by time variations of the high-density
tail of the PDF should be smaller for small $\alpha_{\rm vir,e}$.

\begin{figure}[t]
\includegraphics[width=\columnwidth]{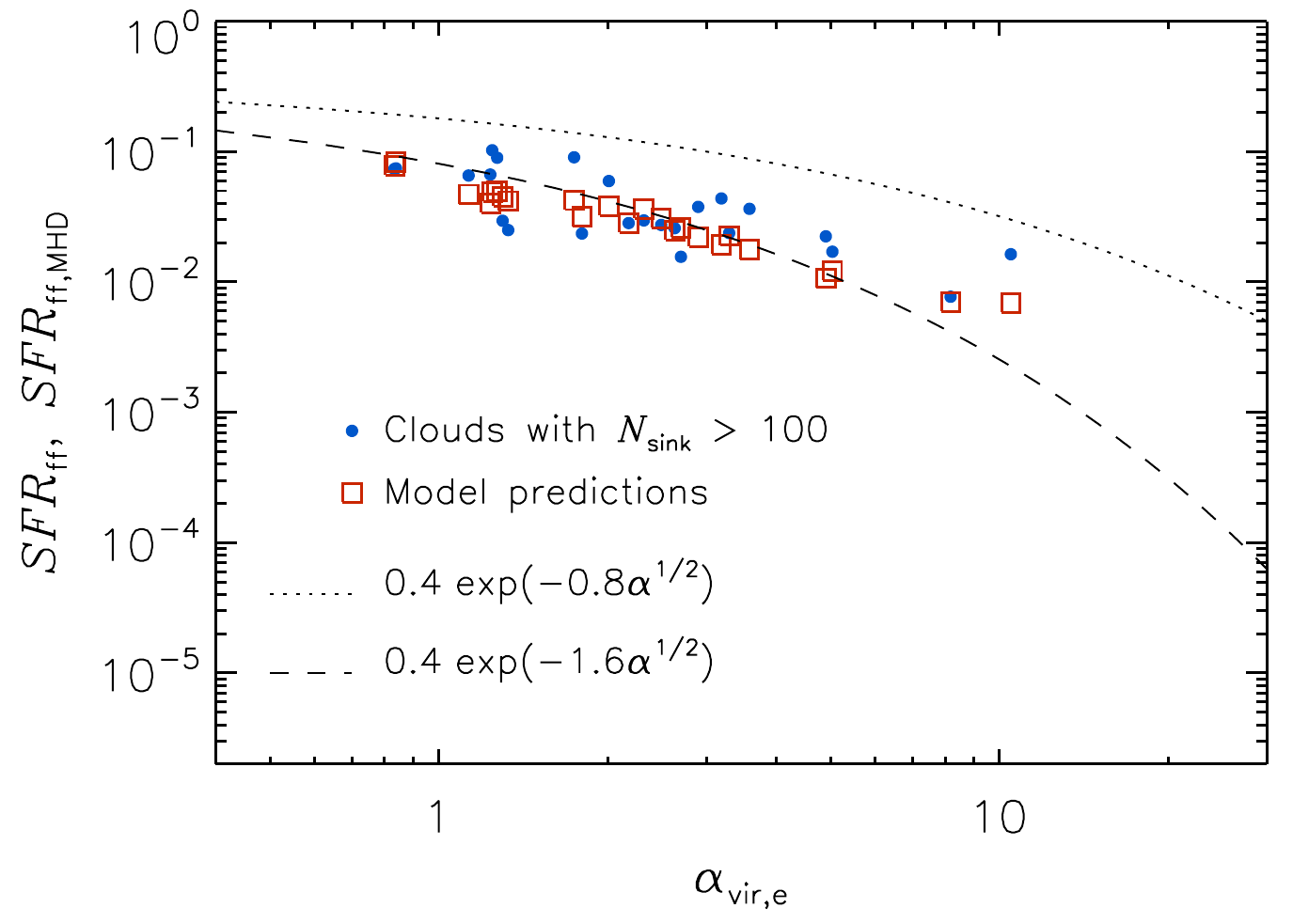}
\caption[]{SFR per free-fall time versus effective virial parameter for a subset of the clouds selected from the simulation (from both cloud catalogs)
harboring more than 100 sink particles (filled circles). The model prediction, $SFR_{\rm ff,MHD}$, for each cloud is also shown (empty squared symbols). 
For reference, we also show the same dashed and dotted curves as in Figures \ref{sfr_alpha} and \ref{sfr_alpha_model}. The scatter in $SFR_{\rm ff}$ 
for this cloud subsample is significantly reduced relative that of the full samples shown in Figures \ref{sfr_alpha}, and the deviations from the model 
predictions are on average less than 50\%.}
\label{sfr_alpha_sinks}
\end{figure}

To evaluate the amplitude of the deviations in the high-density tail of the cloud density PDF relative to the model prediction, we compute, for each cloud,
the mass fraction above the predicted critical density based on the cloud density field from the simulation, $M_{\rm cr}/M_{\rm cl}$, where $M_{\rm cr}$ is the 
mass above the critical density. We then compare this mass fraction with that based on the density PDF predicted by the model for the physical parameters 
of the cloud, $M_{\rm cr,MHD}/M_{\rm cl}$, where $M_{\rm cr,MHD}$ is the mass above the critical density in the theoretical PDF. The comparison is shown
in Figure \ref{sfr_pdf_sfr_mhd} for both cloud catalogs. Because larger values of these mass fractions correspond to higher values of the $SFR_{\rm ff}$,  
Figure \ref{sfr_pdf_sfr_mhd} exhibits the same trend of increasing scatter with decreasing $SFR_{\rm ff}$ as in Figure \ref{sfr_alpha}. The scatter is relatively 
large, though not enough to explain the corresponding scatter in $SFR_{\rm ff}$ in Figure \ref{sfr_alpha}. Furthermore, many clouds still deviate significantly 
from the predicted SFR even if we apply the model with the mass fraction above the critical density directly measured in the clouds, $M_{\rm cr}/M_{\rm cl}$, 
rather than the one predicted by the model, $M_{\rm cr,MHD}/M_{\rm cl}$. Thus, variations of the density PDF cannot be the only explanation for the scatter 
in the relation between $SFR_{\rm ff}$ and $\alpha_{\rm vir,e}$ found in the simulation. 

Random variations in $SFR_{\rm ff}$ due to the finite number of collapsing cores in each cloud certainly contribute to the scatter, particularly in clouds with 
high values of $\alpha_{\rm vir,e}$ that contain a relatively small number of sink particles. This naturally explains the increase in the scatter with increasing 
$\alpha_{\rm vir,e}$ in Figure \ref{sfr_alpha}. We can illustrate the role of low-number statistics by selecting a subsample of clouds containing a large number of
sink particles, $N_{\rm sink}$, which should significantly reduce the effect of random fluctuations of the specific physical realization of the turbulence in each cloud. 
In Figure \ref{sfr_alpha_sinks} we show the case of clouds with $N_{\rm sink}>100$, extracted from both cloud catalogs (filled circles). We also overplot the
model prediction, $SFR_{\rm ff,MHD}$, for each cloud (empty squared symbols). The scatter is very much reduced compared with Figure \ref{sfr_alpha}, and
the values of $SFR_{\rm ff}$ from the simulation deviate from the model predictions by less than 50\% on average. In Figure \ref{sfr_alpha_sinks}, the scatter 
does not show the clear dependence on $\alpha_{\rm vir,e}$ seen in Figure \ref{sfr_alpha}, as we have excluded the clouds with low sink-particle numbers. 
The effect of random variations in $SFR_{\rm ff}$ may be somewhat larger in our simulation than in nature, due to the incompleteness of our stellar IMF below
approximately 5-10 M$_{\odot}$. This will be tested in future works using zoom-in simulations of individual clouds, where the increase resolution will yield a larger
number of collapsing cores and a complete IMF down to a fraction of a solar mass. 

Additional scatter, contributing to the total one in Figure \ref{sfr_alpha}, must arise from the lack of statistical equilibrium in the clouds, due to the specific boundary conditions 
of each cloud and the limited cloud lifetime. The average cloud lifetime is rather short, $t_{\rm life} \approx 4.5\, t_{\rm dyn}$ (see Paper I), and MCs are hardly isolated objects: 
they are part of a complex filamentary network where mass accretion onto the clouds may not be negligible, and the cloud structure is ever changing, with the cloud 
ultimately being dispersed. While the MC turbulence is driven by SNe with an effective outer scale of order 70-100 pc (see Paper I), nearby SNe can also
affect the clouds directly, temporarily changing the relation between $SFR_{\rm ff}$ and $\alpha_{\rm vir,e}$ (for example causing a sudden increase in the cloud rms 
Mach number, while only affecting a relatively small fraction of the cloud mass and thus not modifying the SFR significantly.)

As explained in Section \ref{sec_sfr}, we computed the SFR as an average over a time $\Delta t = 1.68$ Myr. Although the scatter in the SFR may
be reduced by averaging over a significantly longer time (the average dynamical time of the clouds in our samples is nearly 3.0 Myr, so the average lifetime
may be of order 13 Myr), we did not try to define the SFR over larger $\Delta t$. This is because large variations from cloud to cloud are also found 
from the observations, thus it is important to avoid measuring the SFR in the simulation in a way that cannot be related to the observational SFR estimates.
As discussed below, the SFR is measured in nearby MCs by counting the number of protostars, and assuming a typical protostellar lifetime of 2 Myr, 
which is probably a reasonable estimate of the duration of class II protostars \citep{Evans+09,Spezzi+08}. Even when the SFR is measured in a sample of 
clusters selected by free-free emission of ionized gas, the typical cluster lifetime is probably of the order of 2 Myr, at least if the star formation process is 
very rapid, because the free-free emission is expected to strongly decay in a time of order 4 Myr, essentially the lifetime of the massive stars that contribute 
the most to the ionized flux \citep{Murray11}. Thus, our value of $\Delta t$ is comparable to the characteristic age of the young stellar populations used in the 
determination of the SFR in real MCs. Furthermore, as explained in Section \ref{sec_sfr}, we avoid longer $\Delta t$ also to limit uncertainties related to the 
cloud identification, such as the possibility of a significant mass accretion (which we neglect) or the chance that a cloud is completely dispersed during the time 
we measure the SFR.

Given this unavoidable scatter in $SFR_{\rm ff}$ at constant $\alpha_{\rm vir,e}$, we conclude that SFR models should be compared with simulations or 
observations by computing ensemble averages with large cloud samples. This has been achieved with our simulation, allowing us to show that, despite the  
scatter, the values of $SFR_{\rm ff}$ averaged in logarithmic bins of $\alpha_{\rm vir,e}$ closely follow the revised model (see Figure \ref{sfr_alpha}). 
Observational efforts should also aim at compiling very large samples of MCs with estimated values of $SFR_{\rm ff}$ and cloud parameters.

\section{Comparison with Observations} \label{sec_real_sfr}

The average SFR in MCs can be deduced from the global SFR in the Galaxy. For example, \citet{Krumholz+Tan07} derived a value of $SFR_{\rm ff}=0.02$,
and \citet{Murray11} a value approximately three times smaller, $SFR_{\rm ff}=0.006$, due primarily to a different IMF assumption. To move beyond global 
values and derive the dependence of the SFR on physical parameters it is necessary to study the properties of individual MCs and directly measure their SFR. 
Focusing on the most active star-forming regions in the Galaxy, selected by their free-free emission, \citet{Murray11} found a rather large mean SFR, $SFR_{\rm ff}=0.14-0.24$, 
and, more importantly, a very wide range of values, from 0.001 to 0.59. \citet{Vuti+16} selected star-forming regions based on a catalog of HII regions
\citep{Anderson+14}, but derived much lower SFRs, based on the 22 $\mu$m band of the WISE satellite \citep{Wright+10}, with an average value of
$SFR_{\rm ff}=0.007$, of the order of the mean Galactic value. The mid-infrared (MIR) method they used may systematically underestimate the SFR by a factor of 2-3,
so their corrected average value may be as large as 0.02. Although the lists of objects by \citet{Murray11} and \citet{Vuti+16} could in principle be used to compare the SFR 
with the cloud properties, there is a large uncertainty in the association of an HII region with an individual MC. Furthermore, the measured cloud velocity dispersion may be
significantly affected by the feedback from massive stars, especially in the case of the most active star-forming clouds. This would produce a dependence of the virial 
parameter on the SFR, rather than probing the dependence of the SFR on the virial parameter.

\begin{figure}[t]
\includegraphics[width=\columnwidth]{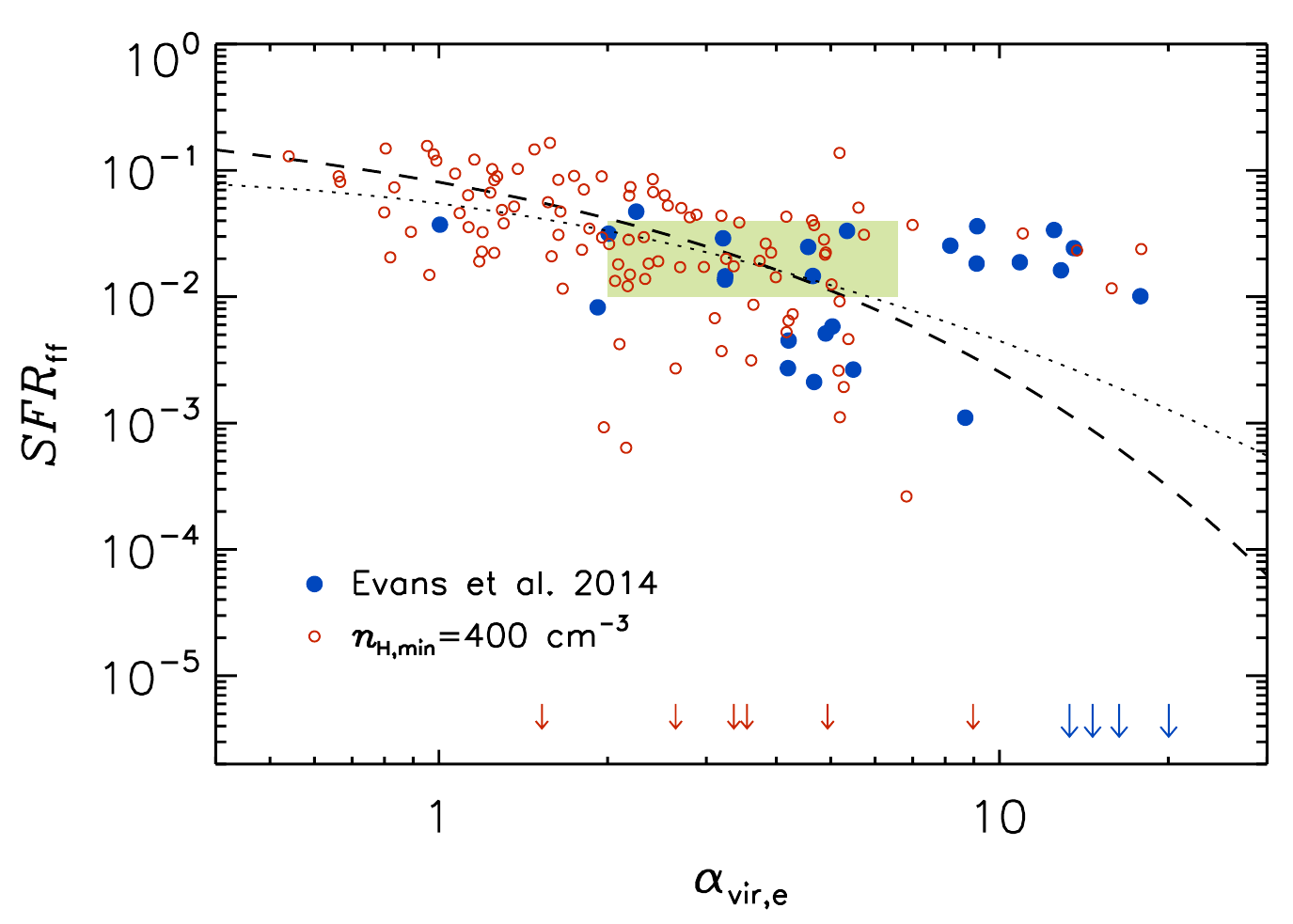}
\caption[]{SFR per free-fall time versus effective virial parameter for the MC sample by \citet{Evans+14} (filled circles) and for the MCs selected from our simulation
with density threshold $n_{\rm H,min}=400$ cm$^{-3}$ (empty circles). The dashed line is the same analytical fit to our model prediction, $SFR_{\rm ff,\alpha}$, as 
in previous figures. The dotted line is our model prediction for a single set of parameters, using the average Mach number value in the observational sample,
${\cal M}=9.4$ (assuming the observed clouds have the same mean temperature as the clouds from the simulation), and the same $b$ and $\beta$ as the average 
values from the simulation, $b=0.48$ and $\beta=0.37$ (because they are unknown for the observed clouds). The green shaded area
shows the estimated $SFR_{\rm ff}$ and $\alpha_{\rm vir,e}$ for the CMZ clouds, from \citet{Federrath+16} and Barnes et al. (2017).}
\label{sfr_alpha_evans}
\end{figure}

To avoid the large uncertainties in the properties of distant clouds associated with HII regions, we compare our numerical results with the observations 
using only data from well-studied nearby MCs. \citet{Evans+14} derived MC properties and SFRs for 29 clouds from the c2d \citep{Evans+03,Evans+09} 
and Gould Belt \citep{Dunham+13} {\it Spitzer} legacy programs. They found an average value of $SFR_{\rm ff}=0.018$ (0.016 including the four clouds 
with $SFR_{\rm ff}=0$), by counting all the protostars in the clouds, and assuming a mean stellar mass of 0.5 M$_\odot$ and a timescale of 2 Myr for 
Class II protostars. 

In Figure \ref{sfr_alpha_evans}, we plot $SFR_{\rm ff}$ versus $\alpha_{\rm vir,e}$ for these nearby MCs (filled circles), and the 
$\alpha_{\rm vir,e}$ values for the four clouds with $SFR_{\rm ff}=0$ (large arrows at $\alpha_{\rm vir,e}>10$). We also plot the clouds from our numerical
catalog with $n_{\rm H,min}=400$ cm$^{-3}$ (empty circles). Including the clouds with $SFR_{\rm ff}=0$, $SFR_{\rm ff}$ decreases with increasing 
$\alpha_{\rm vir,e}$, though not as rapidly as in the simulation. However, the average Mach number of these clouds is 9.4, compared with the value
of 7.7 from our numerical samples (using the same cloud mean temperature in both cases), so the $SFR_{\rm ff}$-$\alpha_{\rm vir,e}$ relation
is expected to be a bit shallower for the observed clouds than for the numerical samples. The dotted line in Figure \ref{sfr_alpha_evans} shows the prediction
of our revised model with ${\cal M}=9.4$, while the dashed line is the analytical fit to the model applied to the clouds from the simulation, as in previous figures.
The number of clouds in the observational sample is too low to define a clear relation between $SFR_{\rm ff}$ and $\alpha_{\rm vir,e}$, but the cloud SFRs 
are approximately distributed around the model prediction shown by the dotted line. Furthermore, the values of $SFR_{\rm ff}$ of the clouds from the simulation 
appear to be consistent with those in nearby MCs. The scatter in $SFR_{\rm ff}$ increases with increasing $\alpha_{\rm vir,e}$ both in the simulation and in the
observations. The nearby MCs with $\alpha_{\rm vir,e}\gtrsim 8$ span a wide range of values, $SFR_{\rm ff}=0-0.04$, approximately the same as in the simulation.  

MCs from the central molecular zone (CMZ) provide an important test for SFR models as well, because they probe star formation under physical conditions
very different from those typical of normal MCs. MCs in the CMZ are much denser and have much larger velocity dispersion for their size than
MCs following standard Larson's relations \citep{Kauffmann+16}. They also have very strong magnetic fields \citep{Pillai+15}. Nevertheless, it turns out that 
the cloud non-dimensional parameters that enter the SFR model are not very different 
from those of normal MCs. Because the properties of the ``Brick" cloud have been carefully derived by \citet{Federrath+16} and thanks to the well defined 
timescale of star formation determined through the orbital model of CMZ clouds by \citet{Kruijssen+15}, the values of $SFR_{\rm ff}$ and $\alpha_{\rm vir,e}$ 
are relatively well determined for the CMZ (Barnes et al. 2017). The SFR value derived by Barnes et al. (2017), $SFR_{\rm ff}=0.01-0.04$, associated with the value 
$\alpha_{\rm vir,e}=4.3\pm2.3$ from \citet{Federrath+16}, are shown by the shaded area in Figure \ref{sfr_alpha_evans}. These values for the CMZ  
are clearly consistent with our numerical results and with the prediction of our revised PN11 model.

%
%
%
%

\section{Summary and Conclusions} \label{sect_conclusions}

We have studied the SFR as a function of cloud parameters by generating a large sample of realistic MCs, formed {\it ab initio} in a simulation
of SN-driven turbulence in an ISM region of 250 pc size. This simulation is a continuation of our previous SN-driven experiment presented in Papers
I, II and III, where we had already demonstrated that we could recover MC properties consistent with the observations. In this work, thanks to a significant 
increase of the spatial resolution and the introduction of sink particles, we can further test if the SFR in the clouds is consistent with the observations as well. 
Although the global SFR in the simulation will be addressed elsewhere, we have anticipated that it corresponds to a depletion time of order 1 Gyr, in agreement 
with global galactic values, which is also an important test for the simulation. The main results of this work are summarized in the following.

\begin{enumerate}

\item The SFR per free-fall time in the MCs selected from the simulation follows a broad probability distribution, with a peak at $SFR_{\rm ff}\approx 0.025$,
and a maximum value of approximately 0.2.

\item On average, $SFR_{\rm ff}$ in the simulation decreases with increasing $\alpha_{\rm vir,e}$.

\item The $SFR_{\rm ff}$-$\alpha_{\rm vir,e}$ relation from the simulation has a large scatter that is not explained by cloud to cloud variations of the other
non-dimensional parameters $b$, $\beta$ and ${\cal M}$. This scatter is most likely due to a combination of time variations in the high-density tail of the gas 
density PDF, random fluctuations of $SFR_{\rm ff}$ in clouds with a low number of stars (sink particles), and a lack of statistical equilibrium of the MC turbulence, 
due to the transient nature of the MCs.

\item The PN11 model has been revised, with the most important modification being the choice of the timescale. While in PN11 we chose the free-fall time
of the critical density, $t_{\rm ff,cr}$, in the revised model we assume it is the timescale of formation of the characteristic post-shock structures responsible
for the high-density tail of the gas density PDF, $t_{\rm PDF}=2\,\theta \, t_{\rm dyn}$, because $t_{\rm ff,cr} < t_{\rm PDF}$. This choice results in a SFR a few times 
lower than in the PN11 model, for characteristic parameters of MCs.

\item Applied to the MCs from the simulation, the revised model results in a $SFR_{\rm ff}$-$\alpha_{\rm vir,e}$ relation with a rather small scatter, which is 
fit well by the relation $SFR_{\rm ff,\alpha} = 0.4 \, \exp(-1.6 \, \alpha_{\rm vir}^{1/2})$. This relation is consistent with our previous result in \citet{Padoan+12sfr}, 
where we had already concluded that $SFR_{\rm ff}$ depends primarily on the virial parameter.

\item The values of $SFR_{\rm ff}$ in the MCs selected from the simulation, averaged in logarithmic intervals of $\alpha_{\rm vir,e}$, follow the prediction 
of our revised PN11 model, $SFR_{\rm ff,\alpha}$ (at least for $\alpha_{\rm vir,e}\lesssim 8$). The model predictions are also followed closely by individual
clouds with $N_{\rm sink}>100$, for which the role of random fluctuations of $SFR_{\rm ff}$ is expected to be small. 

\item The SFR measured in well studied nearby MCs from direct counts of protostars is consistent with the SFR values of the clouds in the simulation 
and with our model predictions. As in the simulation, the scatter is large also in the observational values of $SFR_{\rm ff}$, and increases towards larger
values of $\alpha_{\rm vir,e}$.

\item The SFR in the CMZ, estimated with the aid of a study of the physical properties of the ``Brick" cloud and of an orbital model of the clouds in the CMZ, 
is also consistent with our numerical results and our theoretical predictions.

\end{enumerate}

One of the most valuable findings of this investigation is the large scatter of $SFR_{\rm ff}$, including both large variations from cloud to cloud, even at 
comparable values of $\alpha_{\rm vir,e}$, and time variations within individual clouds. This result has been achieved thanks to the very large number of 
clouds formed {\it ab initio} in the simulation, each contributing to an ensemble of realistic initial and boundary conditions with probability distributions that
may closely match those of real MCs. The instantaneous value of the $SFR_{\rm ff}$ of a given cloud cannot be accurately predicted based on the cloud
physical parameters, and can even deviate significantly from the prediction of the theoretical model, because of the ever-changing nature of MCs embedded
in the complex filamentary structure of the cold ISM, and the continuous driving by SNe, with occasional explosions in close proximity or even within the cloud.
Despite this chaotic nature of the SFR in MCs, we have found that an ensemble average obtained by selecting a very large number of clouds from many snapshots of 
the simulation yields values of $SFR_{\rm ff}$ within logarithmic intervals of $\alpha_{\rm vir,e}$ that nicely fit the model prediction. We conclude that our 
simulation provides support for the theoretical framework of turbulent fragmentation, while exposing the chaotic and unpredictable nature of the star formation process.

\acknowledgements

We thank the anonymous referee for several useful comments and corrections that helped us improve the paper. 
Computing resources for this work were provided by the NASA High-End Computing (HEC) Program through the NASA Advanced 
Supercomputing (NAS) Division at Ames Research Center. PP acknowledges support by the Spanish MINECO under project AYA2014-57134-P. 
TH is supported by a Sapere Aude Starting Grant from The Danish Council for Independent Research. Research at Centre for Star 
and Planet Formation is funded by the Danish National Research Foundation.


\end{document}